\documentclass[useAMS,usenatbib,usegraphicx]{mn2e}
\usepackage{amssymb,graphicx}
\title[Displacement of XRBs from star clusters]
{On the displacement of X-ray binaries from star clusters in star-burst
galaxies}
\author[Zuo \& Li]
{Zhao-Yu Zuo\thanks{E-mail:zuozyu@gmail.com (ZYZ)}, Xiang-Dong
Li\thanks{lixd@nju.edu.cn (XDL)}\\
Department of Astronomy, Nanjing University, Nanjing 210093,
China\\
Key Laboratory of Modern Astronomy and Astrophysics (Nanjing
University), Ministry of Education, Nanjing 210093, China}

\begin{document}
\date{Accepted  . Received  ; in original form }

\pagerange{\pageref{firstpage}--\pageref{lastpage}} \pubyear{2002}

\maketitle

%\label{firstpage}

\begin{abstract}
We have modeled the displacement of luminous X-ray binaries from
star clusters in star-burst galaxies with an evolutionary
population synthesis code developed by \citet{Hurley00,Hurley02}.
In agreement with \citet {kaaret04}, we find significant
spatial offset of X-ray sources
from their parent clusters, and the apparent X-ray luminosity vs. displacement
correlation can be roughly reconstructed. The correlation is not
sensitive to the fundamental properties of the clusters
(e.g., initial mass functions of the binary stars) and the kick velocity
imparted to the newborn compact stars, except the common envelope
parameter $\alpha_{\rm CE}$. We present the distributions of the
main parameters of the current X-ray binaries, which may be used to
constrain the models for the formation and evolution of X-ray
binaries with future optical observations.

\end{abstract}


\section{Introduction}

X-ray binaries (XRBs) contribute a significant fraction of  the
X-ray radiation of normal galaxies \citep{fabb89}. They are binary
systems containing an accreting neutron star (NS) or black hole
(BH) and a normal companion star. Based on the masses $M_{\rm op}$
of the optical companions, XRBs are conventionally divided into
high-mass X-ray binaries (HMXBs) and low-mass X-ray binaries
(LMXBs) \citep[e.g.][]{verbunt94}. In HMXBs, the massive ($M_{\rm
op}\ga 10M_{\odot}$) companions generally have strong stellar
winds (with mass loss rate $\sim 10^{-8}-10^{-5}
M_{\odot}$yr$^{-1}$), part of which can be captured by the compact
star; while LMXBs, in which $M_{\rm op}\la 1.5 M_{\odot}$,
experience mass transfer through
Roche-lobe overflow (RLOF) of the companion, at a rate of $\sim
10^{-10}-10^{-8} M_{\odot}$ yr$^{-1}$. Between them are
intermediate-mass X-ray binaries (IMXBs), in which the companion
stars' masses are in the range $\sim 2 - 10 M_{\odot}$
\citep{heuvel75}. Mass transfer in these binaries also occurs
through RLOF, but on much faster, (sub)thermal timescale of $\sim
10^4-10^5$ yr.

XRBs in galactic disks are thought to have evolved from primordial
binaries, in which a high-mass primary star ($M\ga 10M_{\odot}$)
formed the compact star and the secondary star as its companion.
The formation and evolution of XRBs are often accompanied with
mass transfer and loss of mass and orbital angular momentum
\citep[][for a review]{tauris06}. If the primary star evolves to
be a (super)giant and fills its Roche lobe (RL), its mass is
transferred to the secondary via RLOF. If the mass ratio of the
primary and secondary stars is sufficiently high or the primary
star has a deep convective envelope, the mass transfer process
occurs on a dynamical timescale and is highly unstable, so that a
common-envelope (CE) enshrouding binary results. The secondary
star is captured by the expansion of the giant star and is forced
to move through the giant's envelope. The resulting frictional
drag will cause its orbit to shrink rapidly while, at the same
time, ejecting the envelope before the naked core of the giant
star explodes to form the NS/BH. The binary, if it survives the
supernova (SN), will evolve to be an XRB when the secondary begins
to transfer mass to the compact star. Note that during the
formation and evolution processes of XRBs there may be several
instances of mass transfer and CE phases. For example, HMXBs may
end up in a CE phase, as the NS (or BH) is engulfed by the
extended envelope of its companion. If the system survives after
the CE, an XRB with a Helium companion may be produced. The
formation of LMXBs in globular clusters often invokes dynamical
process such as tidal capture of a low-mass main sequence (MS)
star by a NS \citep{bailyn87} and exchange encounters between an NS
and a primordial binary \citep{davies98}.

Young star clusters and X-ray sources from them have many
interesting aspects on modern astrophysics. Recent data shows that
compact young massive clusters contain a rich population of
massive stars, evidently following a standard Salpeter-like upper
initial mass function \citep[IMF;][]{massey98}. However, there are
also indications that some compact young massive clusters have
either a flatter than normal upper mass function or a cut-off at
low mass \citep{sternberg98,smith01,mcCrady03}. Additionally,
observational studies on the mass ratio of binaries have reached
widely varying, even disparate conclusions \citep[reviewed
in][]{abt83,larson01}. Recent studies find that there are two
populations of secondaries in clusters, which leads to a bimodal
distribution of mass ratios \citep[for more, see \S1.2 in][and
references therein]{kobulnicky07}. For example, \citet{lucy06},
using data from the Ninth Catalogue of Spectroscopic Binary Orbits
\citep{pourbaix04}, reassessed the \citet{hogeveen90} study and
concluded that the data support an excess of $q\simeq 1$ twin
systems. These cases all reveal that potentially more massive
binaries will be produced in these young clusters, which may
present different observational properties when they turn on
X-rays. So investigations on XRBs in young star clusters may help
explore, besides the formation and evolution of XRBs, the
fundamental properties of the clusters as well as recent star
formation processes in galaxies, including stellar population in
galaxies \citep{liu07}, massive star formation and evolution
\citep{Kaper07}.

Using observations from {\it Chandra\/} and NICMOS on board {\it
Hubble Space Telescope\/} ({\it HST\/}), \citet{kaaret04} examined
the spatial offsets between X-ray point sources and star clusters in
three star-burst galaxies. They found that (1) the X-ray sources are
preferentially located near the star clusters, indicating that the
X-ray sources are young objects associated with current star
formation, and (2) brighter X-ray sources preferentially occur
closer to clusters. The displacements of the X-ray sources observed
in these starburst galaxies are likely due to the motion of the X-ray
sources caused by the SN explosions and/or dynamical interactions with
other stars and binaries in the clusters. The absence of bright
X-ray sources with large displacements suggests that there is some
correlation between the luminosity of an X-ray source and its motion
when the X-ray luminosity $L_{\rm X}>10^{38}$ ergs$^{-1}$. They
proposed that these high luminosity sources may be RLOF BH-XRBs
with intermediate mass companions, if emitting
isotropically and running away at a speed of $v\sim$ 10 kms$^{-1}$. If
$v\sim$ 50 kms$^{-1}$ instead, the short lifetime ($\sim$ 4 Myr) of
these luminous sources needs very massive companion stars or
alternatively the luminosity of the sources decreases with age,
which can be examined by an evolutionary population synthesis (EPS)
calculation. They also pointed out that the correlation appears
inconsistent with the highly beamed X-ray emission model
\citep{king01,kording02,kaaret03} because the delay time between the
formation of a BH and the onset of the thermal time-scale mass
transfer is long enough for the X-ray source to move to $\sim 1$ kpc
from its point of origin.

The spatial offset between the X-ray point sources and the star
clusters, especially the X-ray luminosity versus displacement
correlation is determined by the velocity of the binary after the
birth of the NS/BH, the time passed since the SN, and the mass
transfer process. Spatial distribution of X-ray sources in
galactic environment has been investigated both observationally
and theoretically. For example, \citet{paradijs95} and
\citet{white96} have investigated the spatial distribution of NS
and BH LMXBs in our Galaxy, from which they suggested that the
compact objects had received a kick during the SN explosion.
\citet{paczynski90} also studied the spatial distribution of
Galactic NSs in the scope of SN kicks. Recently \citet{zuo08}
modeled the spatial distribution of Galactic XRBs, incorporating
the kinematic evolution of kicked binary systems. \citet {kiel09}
calculated the scale-heights as well as the radial and space
velocity distributions of pulsars, considering their kinematic
evolution within the Galactic potential.
%based on recent works
%by \citet{grimm02,jonker04,temple05,swartz04} and \citet{liujf06}.
However, for XRBs born in star clusters, besides mass ejection
from the binary \citep{nelemans99,heuvel00} and the kick velocity
imparted on the newborn NS caused by the SN explosion
\citep{lyne94}, an ejection speed via dynamical interactions in
clusters \citep{phinney91,kulkarni93,sigurdsson93} should also be
considered. In our calculation we only consider the former two
mechanisms on the motion of XRBs \citep[i.e., primordial
binaries;][]{webbink83,webbink92} though the third one can also
influence the binary population, but not significantly in our
situation, as explained below. Dynamical interactions include
tidal capture, physical collisions and exchange encounters. The
formation rate of NS-XRBs produced through tidal capture process
using Eq.~(3) from \citet{verbunt87} can be estimated as:
%though the third one can also occur on time-scale of a few Myr \citep{portegies99}.
%Here we only estimate the expected number of XRBs produced through
%tidal capture process because the formation of XRBs through
%exchange collisions barely competes with two-body encounters
%\citep{hut83}. Following \citet{verbunt87}, the formation rate in
%clusters of LMXBs through tidal interaction can be expressed as
%follows:
\begin{equation}
R=n_{\rm ns}nv_{\rm rel}\sigma\simeq 6\times10^{-11}\frac{n_{\rm
ns}}{10^2\rm pc^{-3}}\frac{n}{10^4\rm pc^{-3}}
\frac{M_1+M_2}{M_{\odot}}\frac{d}{R_{\odot}}\frac{10\rm
kms^{-1}}{v_{\rm rel}}\rm yr^{-1}\rm pc^{-3}
\end{equation}
where $n_{\rm ns}$ and $n$ are the number densities of NSs and
other stars, respectively, $v_{\rm rel}$ is the relative velocity
between the stars at infinity, $\sigma\simeq\pi
d[2G(M_1+M_2)/v_{\rm rel}^2]$ \citep[i.e., Eq.~(2)
in][]{verbunt87} is the cross section of the two passengers with
small relative velocities, $M_1$ the NS mass, $M_2$ the mass of
the companion star, and $d$ is the distance of closest approach of
the two passengers. Here we adopt $(M_1+M_2)\sim 2M_{\odot}$ and
$d_{\rm max}\sim$ 10$R_{\odot}$ which is estimated roughly from
Eq.~(1) in \citet[][]{verbunt87}. Given the typical values of
young massive clusters as $v_{\rm rel} \sim 10$
 kms$^{-1}$, the core radius $r_{\rm c} \sim 1$ pc, the central
density $\rho_0 \sim 10^4 M_{\odot}$ pc$^{-3}$ \citep{mcCrady03},
we can get $n\sim 2\times 10^4$ pc$^{-3}$, $n_{\rm ns}\sim
0.5f\times 10^2$ pc$^{-3}$ assuming an IMF of \citet{Kroupa},
following the approach (i.e., \S 3.1) of \citet[][]{verbunt87},
here $f\sim 0.2$ is the fraction of NSs remaining in the cluster.
Then we can estimate the predicted number of XRBs in one cluster
is $N_{\rm X}\simeq\frac{4}{3}\pi r_{\rm c}^3R\times T\sim
10^{-2}$ with an X-ray lifetime $T= 10$ Myr. So the expected
number of XRBs produced through this channel in the
\citet{kaaret04} sample is very small ($\ll 1$). Exchange
encounters can also occur on time-scale of a few Myr
\citep{portegies99}, although the formation rate of XRBs through
exchange collisions barely competes with two-body encounters
\citep{hut83}.  However, during the exchange encounters a lower
mass binary star tend to be replaced by a more massive
participant. Hence the effect of exchange encounters is to modify
the mass-ratio distribution of the binaries, making equal-mass
binaries more likely. We examined this effect in our models (i.e.,
models M3 and M7) and found no significant changes.

In the present work, we investigated the kinematic consequences of
XRBs from star clusters in star-burst galaxies from a theoretical
point of view. We used an EPS code to calculate the expected
cumulative distribution of XRB displacements from their parent
clusters. Following the approach of \citet{zuo08}, we calculated
the spatial offset distribution of XRBs with luminosities
$>10^{36}$ ergs$^{-1}$. We mainly examined several parameters,
such as IMF, the secondary star mass function, common envelop
efficiency and kick velocity, which may affect the formation,
evolution and motion of massive XRBs significantly. The objective
of this study is to use the apparent X-ray luminosity versus
displacement correlation to constrain the model parameters and the
fundamental properties of clusters in star-burst galaxies. We also
aim to explore why such correlation exists, which may help
understand the nature of the sources and may be testified by
future observations.

This paper is organized as follows. In \S 2 we describe the
population synthesis method and the input physics for XRBs in our
model. The calculated results are presented in \S 3. Our discussion
and conclusions are in \S 4.

\section{Model}

\subsection{Assumptions and input parameters}

\subsubsection{binary evolution}
To follow the evolution of XRBs, we have used the EPS code
developed by \citet{Hurley00,Hurley02} in our calculations. This
code incorporates evolution of single stars with binary-star
interactions, such as mass transfer, mass accretion, CE evolution,
SN kicks, tidal friction and angular momentum loss mechanics
(i.e., mass loss, magnetic braking and gravitational radiation).
We also updated the original code according to \citet{liu07} and
\citet{zuo08}. The values of other adopted parameters are the same
as the default parameters in Hurley et al. (2002) if not
mentioned.

%For star-burst galaxies, it may undergo a star-burst for several
%million years. Here we adopt a constant star formation rate for 20
In the \citet{kaaret04} samples, the ages of the star clusters
range from 1 to 20 Myr, we then adopt a constant star formation
rate for 20 Myr\footnote{We also extended the star formation
history (SFH) to 50 Myr, and found very small difference in the
final results.}.  For each model, we evolve $10^6$ primordial
systems\footnote{We also varied the number  of the binary systems
by a factor of two, and found no significant difference in the
final results.}, all of which are initially binary systems. We set
up the same grid of initial parameters (primary mass, secondary
mass and orbital separation) as \citet{Hurley02} did and evolve
each binary on the grid. In the following we describe the
assumptions and input parameters in our control model (i.e., Model
M1, listed in Table 1).

\noindent {\em (1) initial parameters}\\
We adopt the IMF of \citet{Kroupa} for the
distribution of the primary mass ($M_1$). For the secondary's mass
($M_2$), we assume a uniform distribution of the mass ratio
$M_2/M_1$ between 0 and 1. A uniform distribution is also taken for
the logarithm of the orbital
separation $\ln a$.

\noindent {\em (2) CE evolution} \\
When mass transfer becomes dynamically unstable, a CE will engulf
the binary. An important parameter in the evolution of close
binaries is the CE parameter $\alpha_{\rm CE}$
\citep{paczynski76,iben93}. It describes the efficiency of
ejecting the envelope ($M_{\rm env}$) by converting orbital energy
(\textbf{$E_{\rm orb}$}) into the kinetic energy that provides the
outward motion of the envelope ($E_{\rm env}$), described as
$E_{\rm env} \equiv \alpha_{\rm CE} \triangle E_{\rm orb}$. There
are several schemes for the CE evolution in the literature
\citep[e.g.,][]{iben93,webbink84,Hurley02}. Here we adopt the
estimation of the reduction of the orbital separation suggested by
\citet{kiel06}
\begin{equation}
\frac{GM_1M_{\rm env}}{\lambda a_{\rm i} r_{\rm L1}}\equiv
\alpha_{\rm CE}[\frac{GM_{\rm c}M_{2}}{2 a_{\rm f}}-\frac{GM_{\rm
c}M_{2}}{2 a_{\rm i}}],
\end{equation}
which yields the ratio of final (post-CE) to initial (pre-CE) orbital
separations as
\begin{equation}
\frac{a_{\rm f}}{a_{\rm i}}= \frac{M_{\rm c}M_{2}}{M_{1}}
\frac{1}{M_{\rm c}M_2/M_1+2M_{\rm env}/(\alpha_{\rm CE} \lambda
r_{\rm L1})}.
\end{equation}
In the above equations the subscripts $f$ and $i$ denote the final and initial
values, respectively; $M_{\rm c}=M_1-M_{\rm env}$ is the core mass
of the primary star ($M_1$) that fills out its RL, $r_{\rm L1}=R_{\rm L1}/a_{\rm i}$ is
the dimensionless RL radius, and $\lambda$ is a
parameter which depends on the stellar mass-density distribution,
respectively. Here
we adopt $\alpha_{\rm CE}=1$ and $\lambda=0.5$. The orbital
separation of the surviving binaries is often reduced by a
factor of $\sim 100$ as a result of the spiral-in. If there is not
enough orbital energy available to eject the envelope, the orbital motion of
the companion during the spiral-in process may be unable to
drive off the envelope of the primary star, resulting
in coalescence rather a compact binary.

\noindent {\em (3) SN kick velocity}\\
When a binary survives a SN
explosion, it receives a velocity kick due to any asymmetry in the
explosion \citep{lyne94}. The kick velocity $v_{\rm k}$ is assumed
to be imparted on the newborn NS with the Maxwellian distribution
\begin{equation}
   P(v_{\rm k})=\sqrt{\frac{2}{\pi}}\frac{v^{2}_{\rm k}}{\sigma_{\rm kick}^{3}}
   \exp(-\frac{v^{2}_{\rm k}}{2\sigma_{\rm kick}^{2}}),
\end{equation}
and we adopt $\sigma_{\rm kick}$=190 kms$^{-1}$ \citep{hansen97}
in our control model. For BH systems, we assume that those BHs
formed with successful SN explosions \citep[the Carbon/oxygen core
mass less than $\sim 7.6 M_{\sun}$,][]{fryer99a,fryer01} were
imparted on a natal kick velocity, which is inversely proportional
to the BH mass, i.e., $v_{\rm kick,BH}=1.4/M_{\rm BH}\times v_{\rm
kick,NS}$. Otherwise no kick velocity is adopted. The velocity
($\textbf{v}_{\rm s}$) of the binary system is related to both the
kick velocity and the orbital velocity of the system, and can be
expressed as \citep[see][for details]{Hurley02},
\begin{equation}
   \textbf{v}_{\rm s}=\frac{M_1^{'}}{M_{\rm b}^{'}}\textbf{v}_{\rm
   k}-\frac{\bigtriangleup M_1M_2}{M_{\rm b}^{'}M_{\rm b}}\textbf{v},
\end{equation}
where $M_{\rm b}=M_1+M_2$ and $M_{\rm
b}^{'}=M_{\rm b}-\Delta M_1$ are the total masses of the system before
and after the SN, respectively;  $M_1^{'}=M_1-\Delta
M_1$ is the current mass of the primary star after losing mass $\Delta M_1$
during the SN, $\textbf{v}$ is the relative orbital velocity of the
stars (expressed as Eq.~A1 in \citet{Hurley02}). Tidal effect is taken into
account to remove any eccentricity induced in a post-SN binary prior to
the onset of mass transfer.

%where $\textbf{v}=-v_{orb}(\sin\beta \textbf{i}}+\cos\beta
%\textbf{j})$ is the relative velocity of the stars (expressed as

%  However we also need to consider parameters which can change the
%evolution of massive X-ray binaries significantly, because it may
%also change the spatial offsets of XRBs from star clusters. For
%star-burst galaxies, the clusters may contain relatively more
%massive stars, and thus

We also construct several other models by changing the key input
parameters (listed in Table 1). Recent observations indicate that
some compact young massive clusters contain relatively more
massive stars \citep{sternberg98,smith01}, we then consider an IMF
from \citet[][i.e., Model M2 and M6]{ballero07}, which is more
skewed towards high mass than in the solar neighbourhood for
comparison. The secondary star mass function can also strongly
affect the formation rate of different type of sources
\citep{kalogera98,belczynski02}. For example, different choice of
the secondary star mass function can change the formation rates of
NS binaries by nearly a factor of $\sim 100$ \citep{fryer99}. Here
we assume the secondary mass following a power-law distribution:
$P(q)\propto q^{\alpha}$, where $q\equiv M_2/M_1$. Most population
synthesis studies adopt a flat mass spectrum (i.e., $\alpha=0$ in
our control model M1) for systems that are likely to interact,
while recent observations seem to be more consistent with ``twins"
being a general feature of the close-binary population
\citep[i.e., Model M3 and M7,][]{dalton95,kobulnicky07}. As stated
before, binary interactions may also tend to modify the mass ratio
to approach unity. So we also adopt $\alpha=1$  to examine its
effect.

Variations in the CE parameter can considerably change the
relative numbers of XRBs. However, reliable value for $\alpha_{\rm
CE}$ has proven difficult to estimate, due to lack of
understanding the complicated processes involved. Generally
$\alpha_{\rm CE} \simeq$ 1 is used but its range can change widely
from $\sim$ 0.1 to $\sim$ 3.0 (see discussion in \S 4). So we also
adopt $\alpha_{\rm CE}=3.0$ for comparison (i.e., models M5, M6,
M7 and M8).

Finally, since the kick velocity $v_{\rm k}$ can affect not only
the global velocity of the binary system but the outcome of the
XRB evolution, we also adopt $\sigma_{\rm kick}=270$ kms$^{-1}$
\citep[i.e., models M4 and M8,][]{Hobbs} for comparison.

\subsubsection{binary motion}
Since star clusters in star-burst regions are usually centrally
concentrated, we assume a spherical potential and adopt the
cylindrical coordinate system ($r$, $\phi$, $z$) centered at the
cluster's center. The potential of a cluster can be constructed
as
\begin{equation}
   \Phi_{\rm N}(r,z)=\frac{-GM}{\sqrt{r^2+z^2}+h},
\end{equation}
where $G$ is the gravitational constant, $h$ the half light radius
and $M$ the total mass of stars within the half light radius. Here
we adopt $M=1.0\times10^{6} M_{\odot}$, and $h=3$ pc
\citep{ho96a,ho96b} in our calculations\footnote{We also adopted
$M=5.0\times10^{5} M_{\odot}$, and found very small difference in
the final results.}. We assume that stars are born uniformly in
the star cluster with  random direction of the initial velocity
vector, which gives the initial velocity vectors $v_{\rm r}$,
$v_{\phi}$, $v_{\rm z}$. Due to the cylindrical symmetry of the
potential, two space coordinates $r$ and $z$ are sufficient to
describe the XRB distributions. Then we integrate the motion
equations \citep[i.e., Eqs.~(19a,b) in][]{paczynski90} with a
fourth-order Runge-Kutta method to calculate the trajectories of
the binary systems and collect the parameters of current XRBs if
turning on X-rays. Finally we project the positions of XRBs on the
$\phi=0$ plane to get the projected distances of XRBs from star
clusters, i.e., $R=((r \cos \varphi)^2+z^2)^{1/2}$ where $\varphi$
is uniformly distributed between 0 and $2\pi$. In our
calculations, the accuracy of integral is set to be $10^{-6}$ and
controlled by the energy integral.

\subsection{X-ray luminosity and source type}

Compact stars in XRBs are powered by either disk or wind
accretion. When a star expands to fill its RL as a result of
stellar evolution, or by angular momentum losses causing
contraction of the orbit, it can transfer masses via an accretion
disk which is fed by RLOF; otherwise, mass-transfer cannot occur
except that there is a stellar wind to power an observable X-ray
source (i.e., wind-accretion). We adopt the same procedures to
calculate X-ray luminosity and divide different types of sources
as in \citet{zuo08} if not mentioned. In particular, for disk-fed sources,
the simulated X-ray luminosity form is as follows:
\begin{eqnarray}
L_{\rm X, 2-10 kev}&=&\left\{
\begin{array} { ll}
  \eta_{\rm bol}\eta_{\rm out}L_{\rm Edd}&\ \rm transients\ in\ outbursts \\
  \eta_{\rm bol}\min(L_{\rm bol},\eta_{\rm Edd}L_{\rm Edd})&\ \rm persistent\
  systems.
\end{array}
\right.
\end{eqnarray}
where $\eta_{\rm bol}=0.3$ is the bolometric correction factor
which is introduced to convert the bolometric luminosity ($L_{\rm
bol}$) to the X-ray luminosity in the $2-10$ keV energy range
\citep{bel03}; $L_{\rm bol}=\eta\dot{M}_{\rm acc}c^2$ where $\eta$
is the efficiency for energy conversion, $\dot{M}_{\rm acc}$ is
the average mass accretion rate for accreting systems and $c$ is
the velocity of light; $\eta_{\rm Edd}$ is the ``Begelman factor"
\citep{rappaport04} to allow super-Eddington luminosities; the
critical Eddington luminosity $L_{\rm Edd} \simeq 4\pi
GM_{1}m_{\rm p}c/\sigma_{\rm T}=1.3 \times
10^{38}(M_{1}/M_{\sun})$\,ergs$^{-1}$ (where $\sigma_{\rm T}$ is
the Thomson cross section, and $m_{\rm p}$ the proton mass). For
transient sources, the luminosities in outbursts are taken to be a
fraction ($\eta_{\rm out}$) of the Eddington luminosity. Here we
adopt $\eta_{\rm out}=0.1$ and 1 for NS systems if the orbital
period $P_{\rm orb}$ is less and longer than 1 day, respectively;
for BH systems, we adopt $\eta_{\rm out}=P_{\rm orb}/24$ hr and
let the maximum peak luminosity not exceed $5L_{\rm Edd}$
\citep{bel03,chen97,Garcia03}. To discriminate transient and
persistent sources, we use the criteria of \citet{paradijs96} for
MS and red-giant donors, and of \citet{ivanova06} for white dwarf
donors, respectively.

\section{Results}

The three star-burst galaxies M82, NGC 1569 and NGC 5253 in
\citet{kaaret04} all contain several young, luminous star clusters
and luminous X-ray sources in them. The number of X-ray sources in
M82/NGC 1569/NGC 5253 is 42/14/10 with the corresponding cluster
number 50/58/13, respectively. The spatial offset between the X-ray
point sources and the star clusters shows that, while the X-ray
sources are generally located near the star clusters, brighter X-ray
sources preferentially occur closer to the clusters, and there is an
absence of very bright sources ($L_{\rm X}>10^{38}$ ergs$^{-1}$) at
relatively large displacements ($>200$ pc) from the clusters. Here
we modeled the kinematic evolution of XRBs from the star clusters.
The calculated results are presented below.

As stated before we constructed 8 models to investigate how the
final results are influenced by the adopted parameters.
Specifically the input parameters in our control model (i.e.,
model M1) are SFH$=20$ Myr, $\sigma_{\rm kick}=190$ kms$^{-1}$,
$\alpha=0$, $\alpha_{\rm CE}=1.0$, and the Kroupa IMF. In other
models we change only one parameter each time, and the model
parameters are listed in Table 1. Figures~1 and 2 show the
simulated cumulative distribution of the X-ray source
displacements (top: Total; middle: NS-XRBs; bottom: BH-XRBs) at
the age of 20 Myr for models M1-M4 and M5-M8, respectively. Note
that we only select sources in the luminosity range
$10^{36}<L_X<10^{38}$ ergs$^{-1}$ in order to compare with
\citet{kaaret04}. We also normalize each histogram by the total
number of XRBs within 1 kpc of a star cluster for each galaxy. We
find that the results are in general agreement with
\citet[][Fig.~2 ]{kaaret04}.

Figures~3 and 4 show the modeled distributions of the X-ray
luminosities ($L_{\rm X}$) at different displacement ($R$) from
the star cluster for models M1-M4 and M5-M8, respectively. The
top, middle, and bottom panels are for all XRBs, NS-XRBs, and
BH-XRBs, respectively. The color bar represents the normalized
number ratio of XRBs in the $R-L_{\rm X}$ plane. Note that NS-XRBs
have relatively lower maximum luminosities than BH-XRBs in all of
the models, due to a lower Eddington accretion rate limit. The
predicted $L_{\rm X}$ vs. $R$ relations in all the models are
roughly compatible with the observations, but differences also
exist. For models M1-M4, there is a scarcity of high-luminosity
($L_X>10^{38}$ ergs$^{-1}$) sources in the $30-100$ pc region,
while there is an overabundance of very luminous sources
($L_X>10^{39}$ ergs$^{-1}$) in the same region for models M5-M8.
In addition, the correlations originate from different stellar
populations in different models. On one hand, models M1-M4 have
similar $L_{\rm X}$ vs. $R$ correlations which are constructed by
both BH-XRBs and NS-XRBs. Note that BH-XRBs dominate at the
high-luminosity ($L_X>10^{38}$ ergs$^{-1}$), small-offset
($10<R<300$ pc) region (named as region A), while NS-XRBs dominate
at the low-luminosity ($L_X< 10^{38}$ ergs$^{-1}$), small-offset
($10<R<300$ pc) region (named as regions B) and large-offset
($300<R<1000$ pc) region (named as region C). On the other hand,
for models M5-M8, the correlations mainly result from BH-XRBs
instead. It is interesting to note that there is a flat-roofed
edge in the $10-100$ pc region, which is not caused by the
Eddington luminosity limit for BH-XRBs.

In order to explore the nature of the XRBs in different regions,
we need to examine their observational properties (i.e., current
mass $M_2$ and spectral type of the donor star, orbital period
$P_{\rm orb}$, and system velocity distribution). We present the
$L_{\rm X}-M_2$ (left), $L_{\rm X}-P_{\rm orb}$ (middle), and
$P_{\rm orb}-M_2$ (right) distributions of XRBs in the $10<R<300$
pc region for models M1-M4 and M5-M8 in Figures~5 and 6,
respectively. Figures~7 and 8 show the distributions of the same
parameters in the region of $300<R<1000$ pc. The detailed source
types in regions A, B and C for different models are also listed
in Tables~2-4, respectively.

Figures~5 and 6 show that, the XRBs in region A are in short
orbital periods ($\sim$ 1 hr), with relatively low-mass ($\sim$ 1-3 $M_{\odot}$)
donors for all models. They are mainly RLOF BH-XRBs
with Helium main-sequence (HeMS) companions (see Table~2). The binary
velocities are $\sim 10-20$ kms$^{-1}$ for models M1-M4,
and $\sim 30-100$ kms$^{-1}$ for models M5-M8.
Considering that they have similar evolutionary timescales
(see Fig.~9 below), the difference in the velocities explains the
different maximum offsets of these high-luminosity sources in different
models. The XRBs in region B  have larger orbital periods
and companion masses. For example, the orbital periods are $\sim 1-20$
hr for models M1 and M3-M8, and can reach $\sim 10^3$ hr for model
M2; the companion masses are $\sim1-4 M_{\odot}$ for models M1 and M4,
$\sim 1-20 M_{\odot}$ for models M5-M8, and can reach  $\sim 60
M_{\odot}$ for models M2 and M3. Table~3 indicates that the majority
($>$ 80$\%$) of these sources
are NS-XRBs. They mainly have HeMS companions
($\sim$ 50$\%$ in model M4 to $\sim$ 85$\%$ in model M7),
with mass transfer either through RLOF or by wind capture.
Figures~7 and 8 show that, the XRBs in region C all have
relatively low-mass ($\sim$ 1-4 $M_{\odot}$) companions. Their
orbital periods are around several hours for all models. The typical
velocities are $\sim 150-300$ kms$^{-1}$, much larger than those
of high-luminosity sources.

We tentatively conclude that the difference in the luminosities for
the sources in regions A, B and C is caused by different types of the
compact objects (BH-XRBs vs. NS-XRBs), of the
donors (HeMS star vs. MS star), and of the accretion modes (RLOF vs.
wind-capture). Sources in region A are mainly BH-XRBs
with RLOF mass transfer from an HeMS companion, hence reaching
very high luminosities. Sources in region B have similar properties
as those in region A, but are mainly
NS-XRBs, with relatively lower luminosities. Sources in region C
are either NS-XRBs (in models M1-M4), or contain a
portion of BH-XRBs with MS companions, so they also cannot
reach luminosities as high as those in region A.

The displacement of the XRBs from the star clusters depends on
their velocities at the moment of the SN explosion and the delay
time from the SN to the beginning of RLOF. In Fig.~9 we present
the distribution of the delay times  for sources in regions A
(dash-dot-dotted line), B (solid line) and C (dotted line),
respectively. We normalize each the histograms by the total number
of X-ray sources in each region. They generally range from 1 to 8
Myr, but the times are a bit shorter in models M1-M4 than in
M5-M8. In the latter they peak at $\sim 5$ Myr.

Note that the $L_{\rm X}$ vs. $R$ relation has different origin
between models M1-M4 ($\alpha_{\rm CE}=1.0$) and models M5-M8
($\alpha_{\rm CE}=3.0$). In the $\alpha_{\rm CE}=1.0$ cases, it is
constructed by both high-luminosity ($L_X>10^{38}$ ergs$^{-1}$)
BH-XRBs and low-luminosity ($L_X<10^{38}$ ergs$^{-1}$) NS-XRBs,
while in the $\alpha_{\rm CE}=3.0$ cases, it results from BH-XRBs
instead, so we need to discuss them separately. In models M1-M4,
luminous BH-XRBs are constrained within $\sim 300$ pc because of
their low velocities, while NS-XRBs, which are relatively dimmer
and faster, can move to region C within $\sim  10$ Myr (see
Table~4). In models M5-M8 , a number of  RLOF BH-XRBs with MS
companions are produced, significantly different from models
M1-M4. These XRBs, occupying region C,  have relatively lower
luminosities and higher velocities compared to the BH-XRBs with
HeMS donors in region A.

%Another remarkable
%difference is that for the $\alpha_{\rm CE}=3.0$ cases, the
%NS-XRBs mainly ($> \sim$ 90$\%$) have MS companions, while only
%$\sim$ 30$\%$ for $\alpha_{\rm CE}=1.0$ cases. The percentage of
%HeMS companions in NS-XRBs are $\sim$ 33-70$\%$ for model M1,M2
%and M4, while $< 5 \%$ for models M5-M8. However for model M3, the
%companions are mainly, $\sim$ 67$\%$ on the naked Helium star
%Hertzsprung Gap state. The remarkable different types of XRB and
%its companion star in this region may provide an interesting clues
%to constrain the common envelope efficiency parameter $\alpha_{\rm
%CE}$.

In order to understand the difference in the binary velocities in
different  models, we examine the distribution of the companion
masses $M_{\rm 2, SNe}$, and orbital periods $P_{\rm orb, SNe}$ at
the moment of SN explosions. It is clearly seen from Eq.~(5) that
the velocity of an XRB is determined by the kick velocity, total
mass and ejected mass, orbital velocity (or orbital period) when
SN occurs. Figures~10 and 11 show the $P_{\rm orb, SNe}-M_{\rm 2,
SNe}$ distribution in regions A, B and C for models M1-M4 and
M5-M8, respectively. High-luminosity sources  in region A (top
panel) generally have longer-period (hence smaller orbital
velocity) and more massive companions, resulting in smaller system
velocity (hence smaller offset from the parent cluster) than
low-luminosity sources (bottom panel).  This fact implies that the
orbital evolution plays an important role in kinematic motion and
the spatial distribution of XRBs.
%
%The
%relatively shorter orbital period of luminous sources (top panel)
%in models M5-M8 also makes a little larger system velocity than
%that of models M1-M4, so it can move to much farther places
%($\sim$ 100 pc) than luminous sources in models M1-M4, as seen in
%Fig.~4.

%
% In addition, a bunch of high-speed, low-luminosity, RLOF
%BH-XRBs with MS companion are found in $300<R<1000$ pc region in
%$\alpha_{\rm CE}=3.0$ cases, while no such kind of BH-XRBs can be
%produced in $\alpha_{\rm CE}=1.0$ cases. Another significant
%difference is the companion type of NS-XRBs in $300<R<1000$ pc
%region. For $\alpha_{\rm CE}=3.0$ cases, they are mainly ($> \sim$
%90$\%$) MS companions while only $\sim$ 30$\%$ are MS companions
%for $\alpha_{\rm CE}=1.0$ cases.

The formation of XRBs usually invokes at least one CE phase. So
the CE  parameter $\alpha_{\rm CE}$ can influence the $L_{\rm X}$
vs. $R$ relation by affecting the binary orbit distribution after
the CE evolution. This further determines the evolutionary state
of the donor star during the XRB phase and the system velocity
after the SN explosion. It has two contrary effects on the
formation and evolution of XRBs. Larger values of $\alpha_{\rm
CE}$ can prevent coalescence of a BH/NS and the companion in a
compact binary during the unstable mass transfer processes, in
favor of the formation of compact XRBs; while for some initially
wide binary systems, a higher value of $\alpha_{\rm CE}$ can cause
the RLOF mass transfer to a NS/BH not to occur within 20 Myr,
decreasing the formation rate of XRBs. So variation of
$\alpha_{\rm CE}$ may lead to different types of XRB populations.
In addition, it can affect the orbital velocity, and the velocity
of the binary. Larger $\alpha_{\rm CE}$ means that more energy was
used to drive off the envelop, leading to wider orbital
separation, and hence smaller orbital velocity after the SN
explosion.

%points.
%The different offset distributions reveal different
%formation and evolution channels for XRBs, so XRB system with
%different secondary mass and spectral type, orbital period, system
%velocity distribution emerges.

To explore the influence of $\alpha_{\rm CE}$ in detail, we
present an example evolutionary sequence for $M_1$, $M_2$, $P_{\rm
orb}$, $L_X$ of an XRB source in region C in Figure~12. We
consider a primordial binary system in a $448.332 \,R_{\odot}$
orbit. The initial stellar masses are 14.309 and 3.833 $M_{\odot}$
for the primary and secondary, respectively. In model M5, the
primary first evolves across the Hertzsprung Gap (HG), expands and
fills its RL at the time 13.7057 Myr, then transfers mass to the
secondary star, which is still on MS. Due to the large mass ratio,
the secondary is unable to accept the overflowing material, and
the system enters the CE stage. The secondary spirals within the
envelope and drives it off on a dynamical timescale, leaving a
compact binary ($a=14.845 R_{\odot}$) with the He core (of mass of
$3.482 M_{\odot}$) of the primary. At the time of $15.7485$ Myr,
the He star evolves across the HG and fills its RL. The system
enters the second CE stage, which makes the orbital separation to
shrink again. In a very short time, the SN explosion occurs and a
NS (of mass  $1.362 M_{\odot}$) is born. The post-SN binary gets a
global velocity of $\sim 200$ kms$^{-1}$ due to the large orbital
velocity before the SN. Now the binary separation becomes $5.395
R_{\odot}$. Because of the small separation, the secondary star
will fill its RL immediately and transfer mass to the NS, leading
to the formation of an NS-XRB. At the time of 16.0062 Myr, the NS
collapses into a BH with mass of $3.008 M_{\odot}$, and a BH-XRB
with an MS companion is formed after twice CE evolutions. The X-ray
luminosity $L_{\rm X}$  first rises to $\sim 10^{39}$ ergs$^{-1}$
and then decreases gradually with time. At the time of 17.69 Myr,
$L_{\rm X}$ decreases to be $\sim 10^{38}$ ergs$^{-1}$ and the
orbital period evolves to be $\sim 10$ hr. By now the binary has
moved for about 2 Myr since the SN explosion, with a displacement
of about 400 pc. The above story explains how BH-XRBs of this type
(RLOF and MS companion) can move to region C. However, when we
decrease the $\alpha_{\rm CE}$ to be 1.0, there isn't sufficient
energy to drive off the entire envelope of the primary star during
the first CE phase, leading to coalescence of the giant core and
the MS star. So no such type of XRBs will be produced.

Note that in the above case, most of the low-luminosity
($L_X<10^{38}$ ergs$^{-1}$) BH-XRBs in region C are formed from
accretion-induced collapse (AIC) of NS systems. Whether AIC of NSs
really happens is still under debate, since quite a few binary
millisecond pulsars, though thought to have experienced extensive
mass accretion, seem to have masses not far from $1.4 M_{\sun}$
\citep[][and references therein]{bassa06}, suggesting significant
mass loss during accretion. So we examine this effect in Figs.~13
and 14, which are the same as Figs.~3 and 4, except that we do not
consider the NS$\rightarrow$BH AIC formation channel. It is
clearly seen that, in the case of $\alpha_{\rm CE}=3.0$ (i.e.,
models M5-M8), the $L_{\rm X}$ vs. $R$ correlation barely exists,
but is constructed by NS-XRBs instead, and BH-XRBs disappear in
region C. We propose that precise velocity measurement of BH-XRBs
may serve as a possible way to discriminate different BH formation
channels.

To explore the evolution of the  $L_{\rm X}$ vs. $R$ relation, we plot
 $L_{\rm X}$ vs. $R$ in Fig.~15  at the time 20, 30, and 40 Myr for
models M1 (top) and M5 (bottom). Note that the correlation in
both cases exists at 20 Myr after the star formation, and
disappears gradually as time goes on. If we shorten the SFH to 5 Myr,
the correlation disappears much earlier, at the time of $\sim$ 25 Myr. So the
$L_{\rm X}$ vs. $R$ correlation cannot hold all the time for
individual cluster.

\section{Discussion and Summary}

We have used an EPS code to calculate the spatial offset
distribution of XRBs from their parent star cluster in star-burst
galaxies. We used the apparent X-ray luminosity versus
displacement correlation to constrain models of XRBs. Our study
shows that the correlation can be roughly reproduced with all
models considered, but significant differences exist when changing
the common envelope parameter $\alpha_{\rm CE}$. In the
$\alpha_{\rm CE}=1.0$ cases (models M1-M4), the $L_{\rm X}$ vs.
$R$ relation is constructed by both high-luminosity ($L_X>10^{38}$
ergs$^{-1}$), small-offset ($10<R<100$ pc) BH-XRBs and
low-luminosity ($L_X<10^{38}$ ergs$^{-1}$), large-offset
($300<R<1000$ pc) NS-XRBs, while for $\alpha_{\rm CE}=3.0$ (models
M5-M8), it is mainly determined by BH-XRBs. In this case, a bunch
of high-speed ($\sim$ 150 kms$^{-1}$), low-luminosity, RLOF XRBs
with MS donors dominate the sources in the $300<R<1000$ pc region,
which are not seen when  $\alpha_{\rm CE}=1.0$. The detailed
source types in region A, B and C are listed in Tables~2-4, and
summarized below.

(1) The high-luminosity ($L_X>10^{38}$ ergs$^{-1}$) sources in
the $10<R<300$ pc region in all models are mainly RLOF BH-XRBs
with HeMS companions. They all have short orbital period ($\sim$ 1
hr) and low-mass ($\sim$ 1-3 $M_{\odot}$) companions. The system
velocity is $\sim$ 10-20 kms$^{-1}$ for models M1-M4, and slightly
higher, $\sim$ 30-100 kms$^{-1}$ for models M5-M8.

(2) The low-luminosity ($L_X<10^{38}$ ergs$^{-1}$) sources in
the $10<R<300$ pc regions in all models are mainly ($>$ 80$\%$)
NS-XRBs with HeMS companions (from $\sim$ 50$\%$ in model M4 to $\sim$
85$\%$ in model M7), transferring mass through RLOF or
wind-capture. The orbital periods are  $\sim 1-20$ hr for models M1 and
M3-M8, and can reach $\sim 10^3$ hr for model M2. The companion
masses are  $\sim 1-4 M_{\odot}$ for models M1 and M4, $\sim 1-20
M_{\odot}$ for models M5-M8 and can reach $\sim 60 M_{\odot}$
for models M2 and M3.

(3) The XRBs in the $300<R<1000$ pc regions in all models all have
relatively low-mass ($\sim$ 1-4 $M_{\odot}$) companions. Their
orbital periods are about several hours, and their system velocities
are $\sim$ 150-300 kms$^{-1}$. For models M1-M4, only NS-XRBs can be
found in this region, with both MS and HeMS donors. For models
M5-M8, there are nearly comparable number of BH-XRBs and NS-XRBs,
with mainly MS donors. The BH-XRBs in these models are from AIC of
NS systems.

Our results indicate that the IMFs of the primary and  the secondary stars, and
the kick velocity can affect the formation and evolution
of massive XRBs, but not so significant to
constrain models by comparing with the observed $L_{\rm X}$ vs. $R$ relation.
However, changing the CE parameter $\alpha_{\rm CE}$ can make significant
differences in stellar components that construct the
correlation. We suggest the remarkable different types of XRB in the
$300<R<1000$ pc region may provide an
interesting clue to constrain the CE parameter.

Common envelope evolution has long been the subjects of studies, from
early semi-empirical estimates by \citet{ostriker75} and \citet{paczynski76},
followed by series of works by, e.g., \citet{webbink92}, \citet{sandquist00},
and \citet{taam00}, to recent 3-D hydrodynamical simulations
by \citet{ricker08}.
However, reliable value of $\alpha_{\rm CE}$ is still uncertain due
to lack of understanding of the processes involved. Previous
semi-empirical estimates  suggested that
$\alpha_{\rm CE}$ is in the range 0.6-1.0 \citep{iben89,tutukov93}
and 0.3-0.6 \citep{livio88,Taam89}, but the value was later suggested to be
greater
than unity if additional energy source is considered, such as
thermal and ionization energy of the envelope \citep{han95} or
possibly nuclear energy generated in the region close to the
in-spiraling companion \citep{taam94} and others \citep[see][for
more]{iben93}. In some evolutionary calculations, high value of $\alpha_{\rm CE}$
($\sim$ 3) was required in order to model the formation of
low-mass BH binaries \citep{portegies97}, low- and intermediate-mass
BH-XRBs \citep{kalogera99} and low-mass, short-period BH
binaries \citep{podsiadlowski03}. Note in our calculations, much
more BH-XRBs can be produced in the $\alpha_{\rm CE}=3.0$ cases than
in the $\alpha_{\rm CE}=1.0$ cases, which is consistent with their
findings.

Our results are subject to some uncertainties and simplified
treatments. For example, in our calculations, we only consider the
primordial binaries. However, dynamical interactions in star
clusters can produce new XRBs and modify the IMFs of the
primordial binary stars, though the proportion may be small, as
estimated above. Another simplified treatment is taken for the CE
evolution. Besides $\alpha_{\rm CE}$, the $\lambda$-parameter,
which describes the binding energy between the envelope and the
core of the donor star, can also affect the CE evolution. It
depends on the structure of the stellar envelope, and consequently
on the evolutionary stage of the star. In our calculation, we
adopt a constant (i.e., $\lambda=0.5$) disregarding the evolution
of the donor star at the onset of the mass transfer process,
though the value of $\lambda$ changes remarkably. For example,
typical value of $\lambda$ was found to be between 0.2 and 0.8,
but can be greater than 5 for the asymptotic giant branch of
lower-mass stars \citep{dewi00}. Recent work by \citet{xu10}
showed that for more massive stars (i.e., $>10 M_{\odot}$), the
value of $\lambda$ may be low as 0.1. So incorporating $\lambda$
as a function of stellar radius will help model the $L_{\rm X}$
vs. $R$ relation more precisely, and put more realistic
constraints on the model parameters, though it is beyond the scope
of this paper.

% Our calculations show that with different $\alpha_{\rm CE}$
%value, the XRB may experience different formation and evolution
%channel, so will have different orbital period, secondary mass and
%spectral type, system velocity and so on. Note that when
%$\alpha_{\rm CE}=3.0$, the high luminosity ( $L_{\rm X}>10^{38}
%\rm ergs^{-1}$) sources which construct the correlation are mainly
%low velocity (about 30-100 kms$^{-1}$) RLOF BH-XRB systems with
%low mass HeMS companions, while low luminosity ( $L_{\rm
%X}<10^{38} \rm ergs^{-1}$) sources are mainly high speed RLOF
%BH-XRBs, which is formed from AIC of NS systems, and their
%companions are mainly MS stars. When we decrease the value of
%$\alpha_{\rm CE}$, BH-XRBs are all RLOF XRB systems with low mass
%HeMS companions. Their system velocity is low, about 10-20
%kms$^{-1}$ so they can not move too far. And the type of
%companions in $300<R<1000$ pc region is also different remarkably
%from $\alpha_{\rm CE}=3.0$ cases. It reveals that more
%observations on the orbital parameters and companion type may help
%us understand the formation and evolution, especially the
%evolution of CE of these sources. Our work motivates further
%effects to explore the spatial offset of luminous XRBs in star
%clusters.

\section{Acknowledgments}

We are grateful to the referee for his helpful comments and suggestions
that greatly improved the paper. We also thank Xi-wei
Liu and Xiao-jie Xu for useful discussions. This work was
supported by the Natural Science Foundation of China (under grant
number 10873008) and the National Basic Research Program of China
(973 Program 2009CB824800). ZYZ was also supported by the Jiangsu
Project Innovation for PhD candidates (0201001504).
%We are also very grateful to an anonymous referee
%whose comments and suggestions largely improved the clarity of
%this paper.
%This work was supported by the Natural Science Foundation of China
%under grant numbers 10573010 and 10221001.

%
\clearpage

\begin{table}
\caption{Parameters adopted for each model. Here $\alpha_{\rm CE}$
is the CE parameter, $q$  the initial mass ratio, $\sigma_{\rm
kick}$ the dispersion of kick speed, IMF is the initial mass
function.} \centering
\begin{tabular}{ccccc}\hline\hline
   Model & $\alpha_{\rm CE}$ & P(q)    & $\sigma_{\rm kick}$ & IMF \\
          &              &   & km/s & \\ \hline
      M1 & 1.0 & $\propto q^{0}$ & 190      &   Kroupa\\
      M2 & 1.0 & $\propto q^{0}$ & 190      &   Ballero\\
      M3 & 1.0 & $\propto q^{1}$ & 190      &   Kroupa\\
      M4 & 1.0 & $\propto q^{0}$ & 270      &   Kroupa\\
      M5 & 3.0 & $\propto q^{0}$ & 190      &   Kroupa\\
      M6 & 3.0 & $\propto q^{0}$ & 190      &   Ballero\\
      M7 & 3.0 & $\propto q^{1}$ & 190      &   Kroupa\\
      M8 & 3.0 & $\propto q^{0}$ & 270      &   Kroupa\\\hline
\end{tabular}
\end{table}

%\begin{table}
%\caption{The detailed source types in $10<R<300$ pc regions. Here
%'BHRLO' represents RLOF BH-XRB systems. 'BHMS' represents BH-XRBs
%with MS companion.} \centering
%\begin{tabular}{ccccccccc}\hline\hline
%   Model & BH$\%$ & $\frac{\rm BHRLO}{\rm BH}$  & $\frac{\rm NSRLO}{\rm NS}$ & $\frac{\rm N(>10^{38} erg/s)}{\rm N(>10^{36} erg/s)}$ & $\frac{\rm BHMS}{\rm BH}$ & $\frac{\rm BHHeMS}{\rm BH}$
%   & $\frac{\rm NSMS}{\rm NS}$ & $\frac{\rm NSHeMS}{\rm NS}$ \\ \hline
%      M1 & 27 & 88 & 28 & 24 & 8  & 90 & 11 & 80\\
%      M2 & 68 & 82 & 59 & 56 & 15 & 82 & 21 & 56\\
%      M3 & 42 & 93 & 63 & 46 & 5  & 93 & 7  & 79\\
%      M4 & 39 & 94 & 62 & 48 & 4  & 94 & 4  & 89\\
%      M5 & 69 & 97 & 60 & 64 & 11 & 85 & 36 & 61\\
%      M6 & 72 & 95 & 60 & 68 & 8  & 86 & 31 & 65\\
%      M7 & 69 & 96 & 48 & 67 & 2  & 93 & 13 & 82\\
%      M8 & 62 & 95 & 59 & 60 & 7  & 90 & 36 & 60\\\hline
%\end{tabular}
%\end{table}

\begin{table}
\caption{The detailed types of sources in region A ($L_{\rm
X}>10^{38}$ ergs$^{-1}$, $10<R<300$ pc). Here ``BH(NS)RLO",
``BH(NS)MS" and ``BH(NS)HeMS" represent RLOF BH(NS)-XRBs,
BH(NS)-XRBs with MS companions and BH(NS)-XRBs with HeMS
companions, respectively. BH$\%$ represents the percentage of
BH-XRBs in region A, $\frac{\rm N(>10^{38} erg/s)}{\rm N(>10^{36}
erg/s)}$ represents the percentage of high-luminosity ($L_{\rm
X}>10^{38}$ ergs$^{-1}$) sources in $10<R<300$ pc region.}
\centering
\begin{tabular}{ccccccccc}\hline\hline
   Model & BH$\%$ & $\frac{\rm BHRLO}{\rm BH}$  & $\frac{\rm NSRLO}{\rm NS}$ & $\frac{\rm N(>10^{38} erg/s)}{\rm N(>10^{36} erg/s)}$ & $\frac{\rm BHMS}{\rm BH}$ & $\frac{\rm BHHeMS}{\rm BH}$
   & $\frac{\rm NSMS}{\rm NS}$ & $\frac{\rm NSHeMS}{\rm NS}$ \\ \hline
      M1 & 99 & 99 & 100 & 22 & 1  & 97 & 0  & 100\\
      M2 & 85 & 99 & 100 & 49 & 11 & 88 & 0  & 100\\
      M3 & 82 & 99 & 100 & 45 & 0  & 99 & 0  & 100\\
      M4 & 99 & 96 & 100 & 24 & 11 & 84 & 0  & 100\\
      M5 & 97 & 99 & 100 & 64 & 9  & 88 & 95 & 5\\
      M6 & 97 & 99 & 100 & 67 & 6  & 90 & 91 & 9\\
      M7 & 97 & 99 & 100 & 68 & 1  & 95 & 43 & 57\\
      M8 & 98 & 99 & 100 & 60 & 5  & 92 & 96 & 4\\\hline
\end{tabular}
\end{table}

\begin{table}
\caption{Same as in Table~2 but for sources in region B
($10^{36}<L_{\rm X}<10^{38}$ ergs$^{-1}$, $10<R<300$ pc). Here
$\frac{\rm N(10^{36}<L_{\rm X}<10^{38} erg/s)}{\rm N(>10^{36}
erg/s)}$ represents the percentage of low-luminosity
($10^{36}<L_{\rm X}<10^{38}$ ergs$^{-1}$) sources in $10<R<300$ pc
region.} \centering
\begin{tabular}{ccccccccc}\hline\hline
   Model & BH$\%$ & $\frac{\rm BHRLO}{\rm BH}$  & $\frac{\rm NSRLO}{\rm NS}$ & $\frac{\rm N(10^{36}<L_{\rm X}<10^{38} erg/s)}{\rm N(>10^{36} erg/s)}$ & $\frac{\rm BHMS}{\rm BH}$ & $\frac{\rm BHHeMS}{\rm BH}$
   & $\frac{\rm NSMS}{\rm NS}$ & $\frac{\rm NSHeMS}{\rm NS}$ \\ \hline
      M1 & 3  & 2  & 28 & 78 & 70  & 28 & 11 & 80\\
      M2 & 11 & 0  & 58 & 51 & 80  & 18 & 23 & 70\\
      M3 & 8  & 34 & 58 & 55 & 52  & 41 & 8  & 76\\
      M4 & 7  & 19 & 51 & 76 & 62  & 20 & 21 & 47\\
      M5 & 13 & 63 & 57 & 36 & 37  & 38 & 33 & 64\\
      M6 & 20 & 46 & 57 & 33 & 29  & 48 & 27 & 70\\
      M7 & 11 & 29 & 44 & 32 & 13  & 64 & 11 & 84\\
      M8 & 9  & 34 & 57 & 40 & 38  & 56 & 35 & 61\\\hline
\end{tabular}
\end{table}

\begin{table}
\caption{Same as in Table~2 but for sources in region C ($L_{\rm
X}>10^{36}$ ergs$^{-1}$, $300<R<1000$ pc). $\frac{\rm N(>10^{38}
erg/s)}{\rm N(>10^{36} erg/s)}$ represents the percentage of
high-luminosity ($L_{\rm X}>10^{38}$ ergs$^{-1}$) sources in
region C.}
\begin{center}
\begin{tabular}{ccccccccc}\hline\hline
   Model & BH$\%$ & $\frac{\rm BHRLO}{\rm BH}$  & $\frac{\rm NSRLO}{\rm NS}$ & $\frac{\rm N(>10^{38} erg/s)}{\rm N(>10^{36} erg/s)}$ & $\frac{\rm BHMS}{\rm BH}$ & $\frac{\rm BHHeMS}{\rm BH}$
   & $\frac{\rm NSMS}{\rm NS}$ & $\frac{\rm NSHeMS}{\rm NS}$ \\ \hline
      M1 & 0  & 0   & 82 & 0  & 0  & 0    & 37 & 44\\
      M2 & 0  & 0   & 23 & 0  & 0  & 0    & 15 & 69 \\
      M3 & 0  & 0   & 33 & 0  & 0  & 0    & 26 & 0$^a$ \\
      M4 & 0  & 0   & 60 & 0  & 0  & 0    & 21 & 33\\
      M5 & 38 & 100 & 95 & 14 & 95 & 5    & 90 & 4 \\
      M6 & 59 & 100 & 99 & 19 & 96 & 4    & 95 & 1 \\
      M7 & 35 & 100 & 99 & 21 & 71 & 29   & 98 & 0 \\
      M8 & 40 & 100 & 96 & 9  & 99 & 1    & 88 & 2 \\\hline
\end{tabular}\\
\end{center}
%\raggedright \\
\noindent $^a$ naked Helium star Hertzsprung Gap: 67$\%$ \\
%$^b$: Hertzsprung Gap: 21$\%$ \\
\end{table}

\clearpage

\begin{figure}
  \centering
   \includegraphics[width=0.8\linewidth]{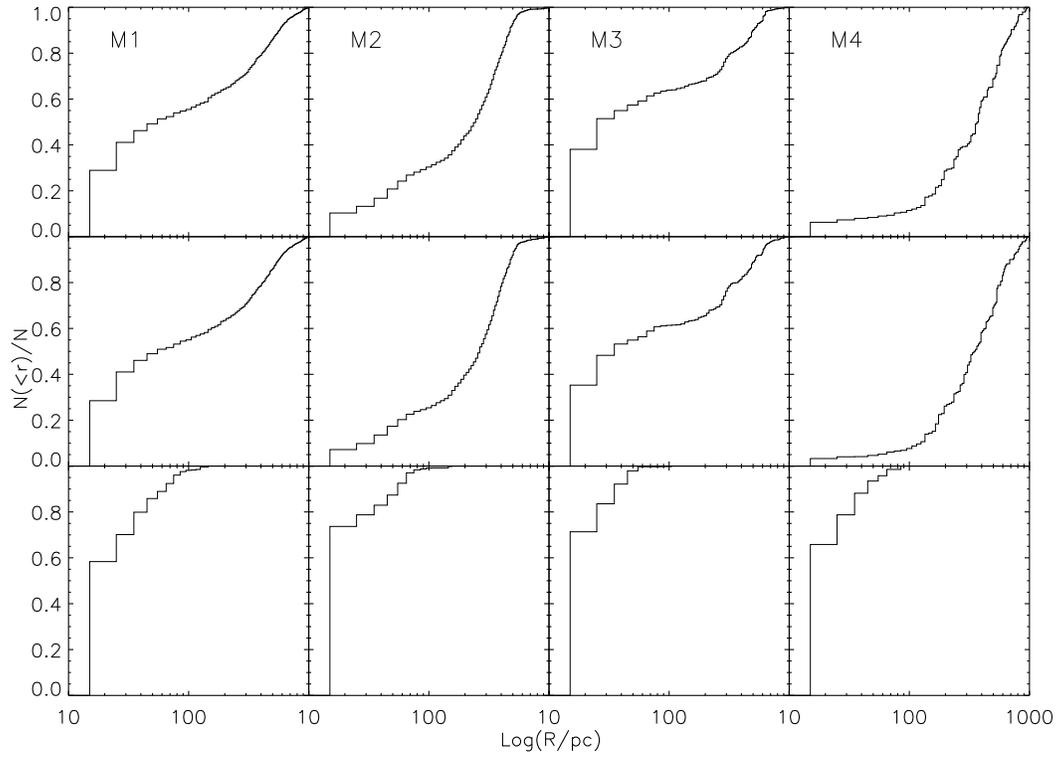}
\caption{The normalized cumulative distribution for the numbers of
ALL-XRBs (top), NS-XRBs (middle) and BH-XRBs (bottom), respectively.
From left to right are models M1-M4, respectively.}
  \label{Fig. 1a}
\end{figure}

\begin{figure}
  \centering
   \includegraphics[width=0.8\linewidth]{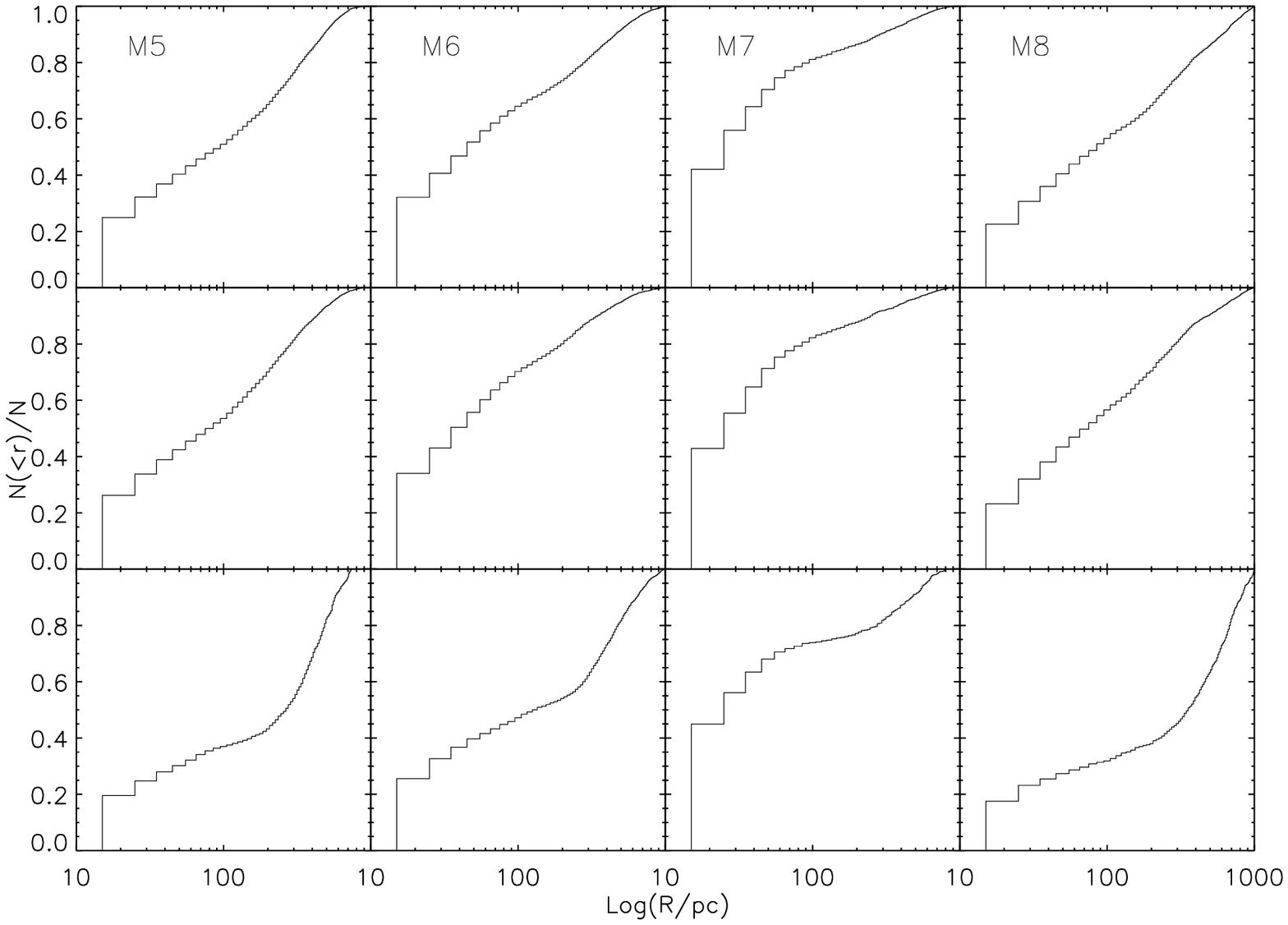}
\caption{Same as Fig.~1 but for sources in models M5-M8 from left to
right, respectively.}
  \label{Fig. 1b}
\end{figure}

\begin{figure}
  \centering
   \includegraphics[width=0.8\linewidth]{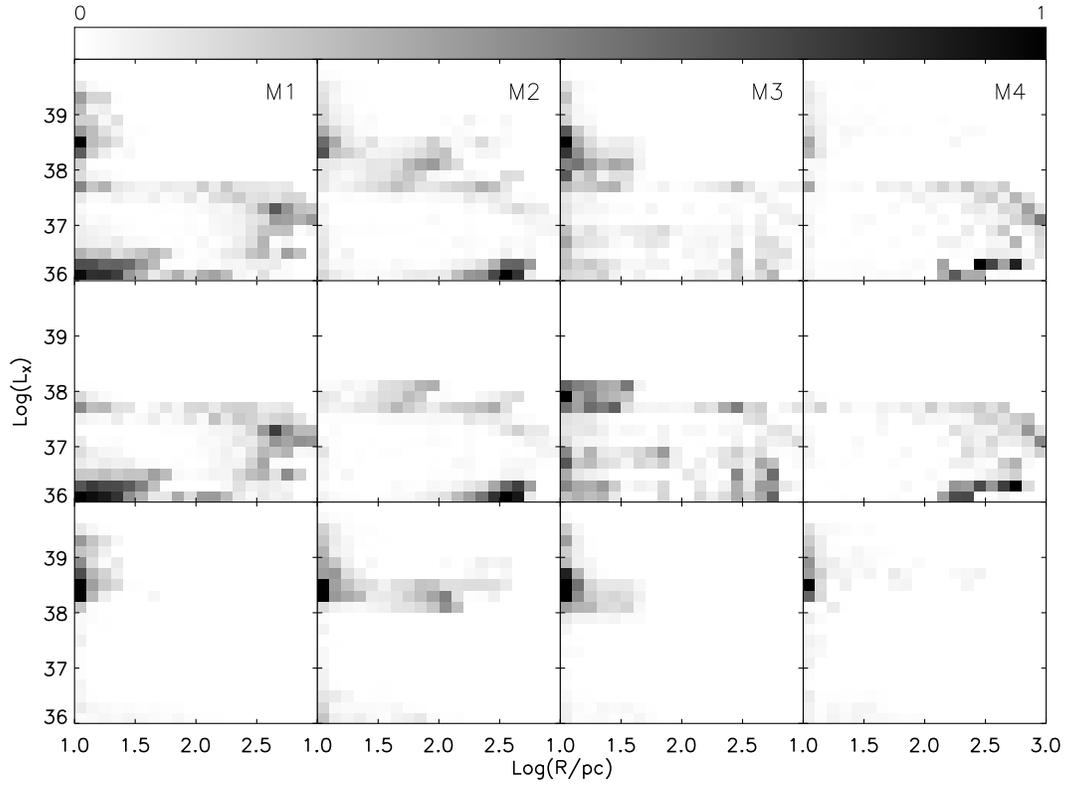}
\caption{The $L_{\rm X}-R$ distribution for ALL-XRBs (top), NS-XRBs
(middle) and BH-XRBs (bottom), respectively. From left to right are
models M1-M4, respectively.}
  \label{Fig. 2a}
\end{figure}

\begin{figure}
  \centering
   \includegraphics[width=0.8\linewidth]{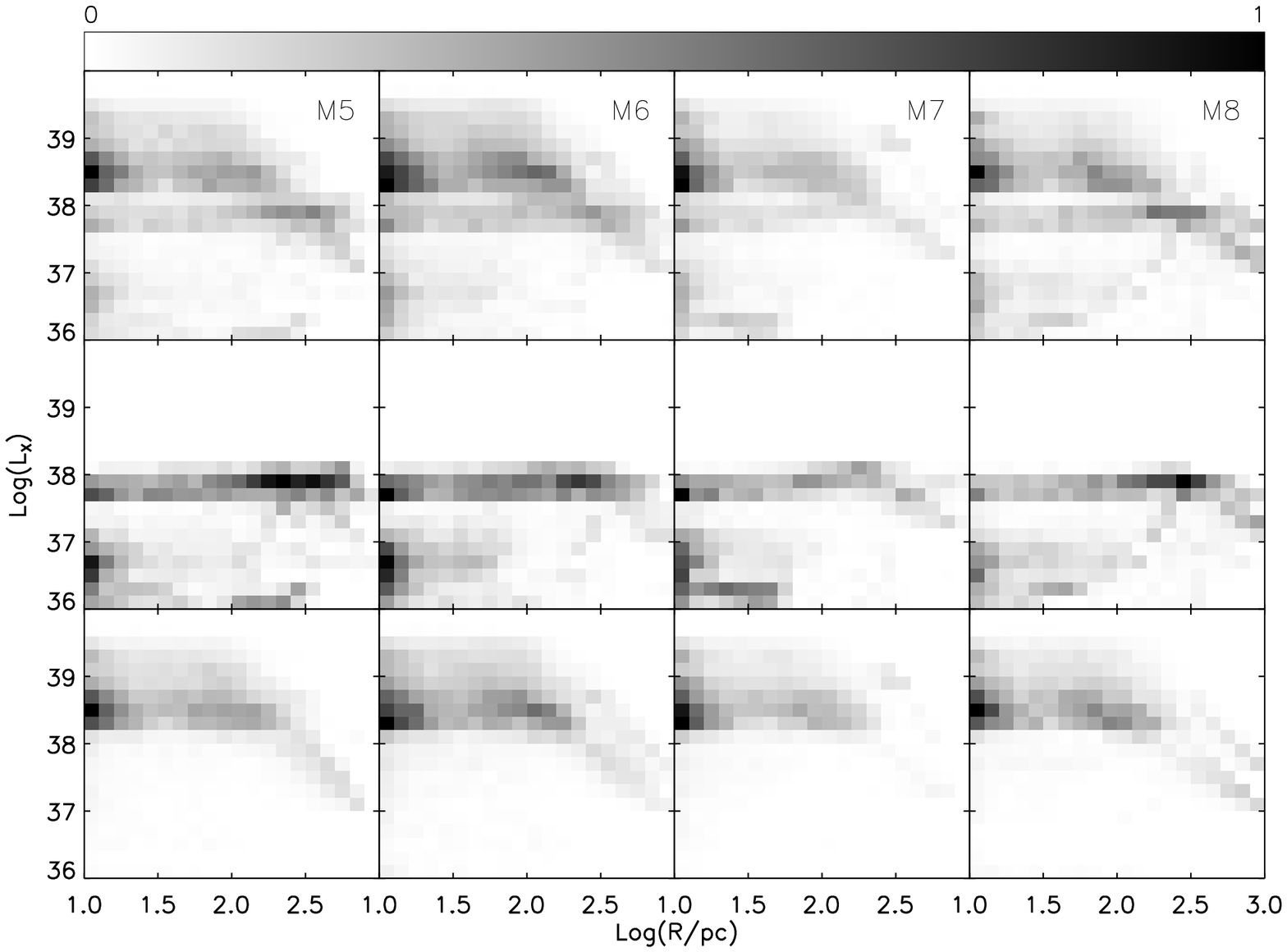}
\caption{Same as Fig.~3 but for sources in models M5-M8 from left to
right, respectively.}
  \label{Fig. 2b}
\end{figure}

\begin{figure}
  \centering
   \includegraphics[width=0.8\linewidth]{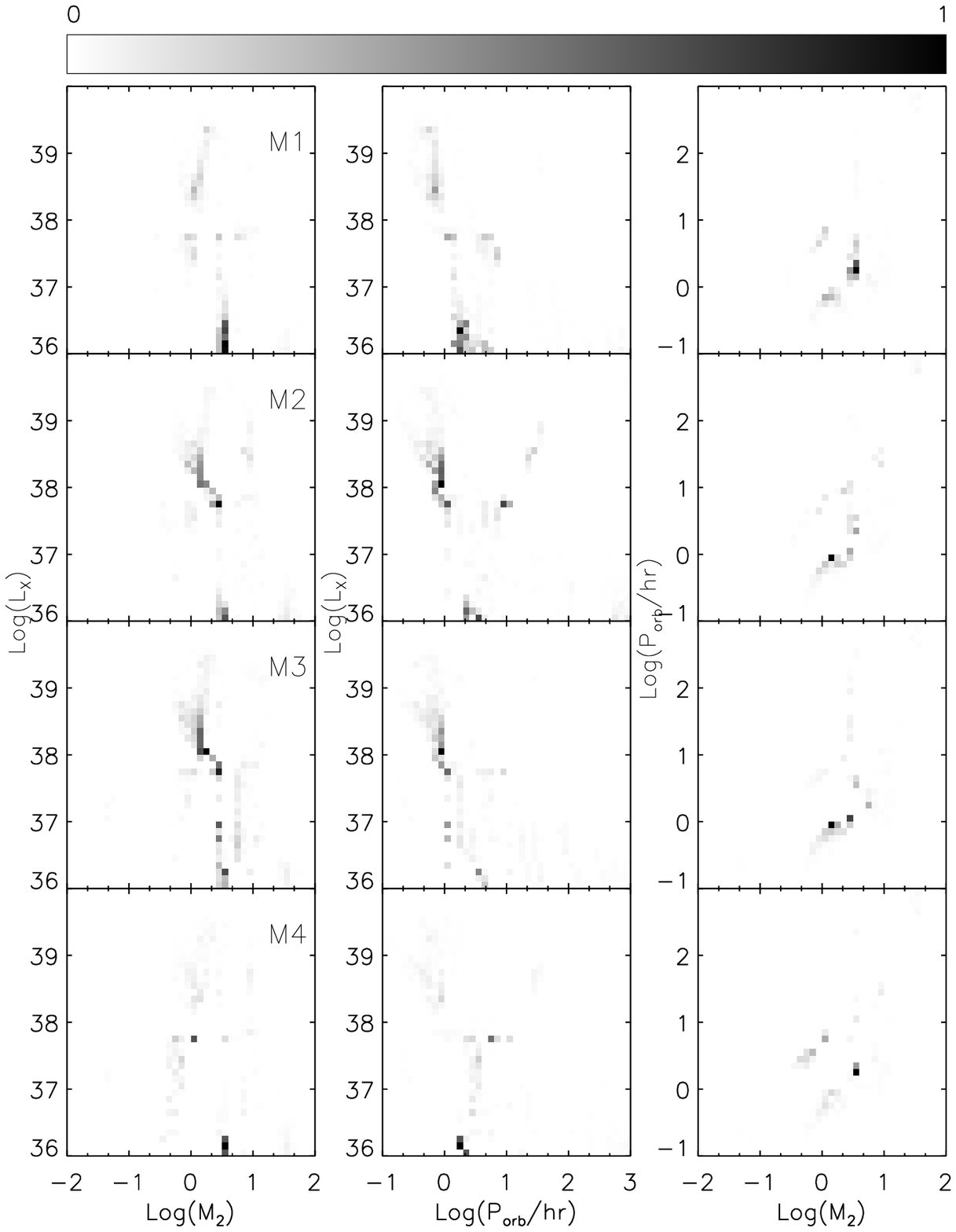}
\caption{The $L_{\rm X}-M_2$, $L_{\rm X}-P_{\rm orb}$, and $P_{\rm
orb}-M_2$ distributions in the $10<R<300$ pc region from left to
right, respectively. From top to bottom are models M1-M4,
respectively.}
  \label{Fig. 3a}
\end{figure}

\begin{figure}
  \centering
   \includegraphics[width=0.8\linewidth]{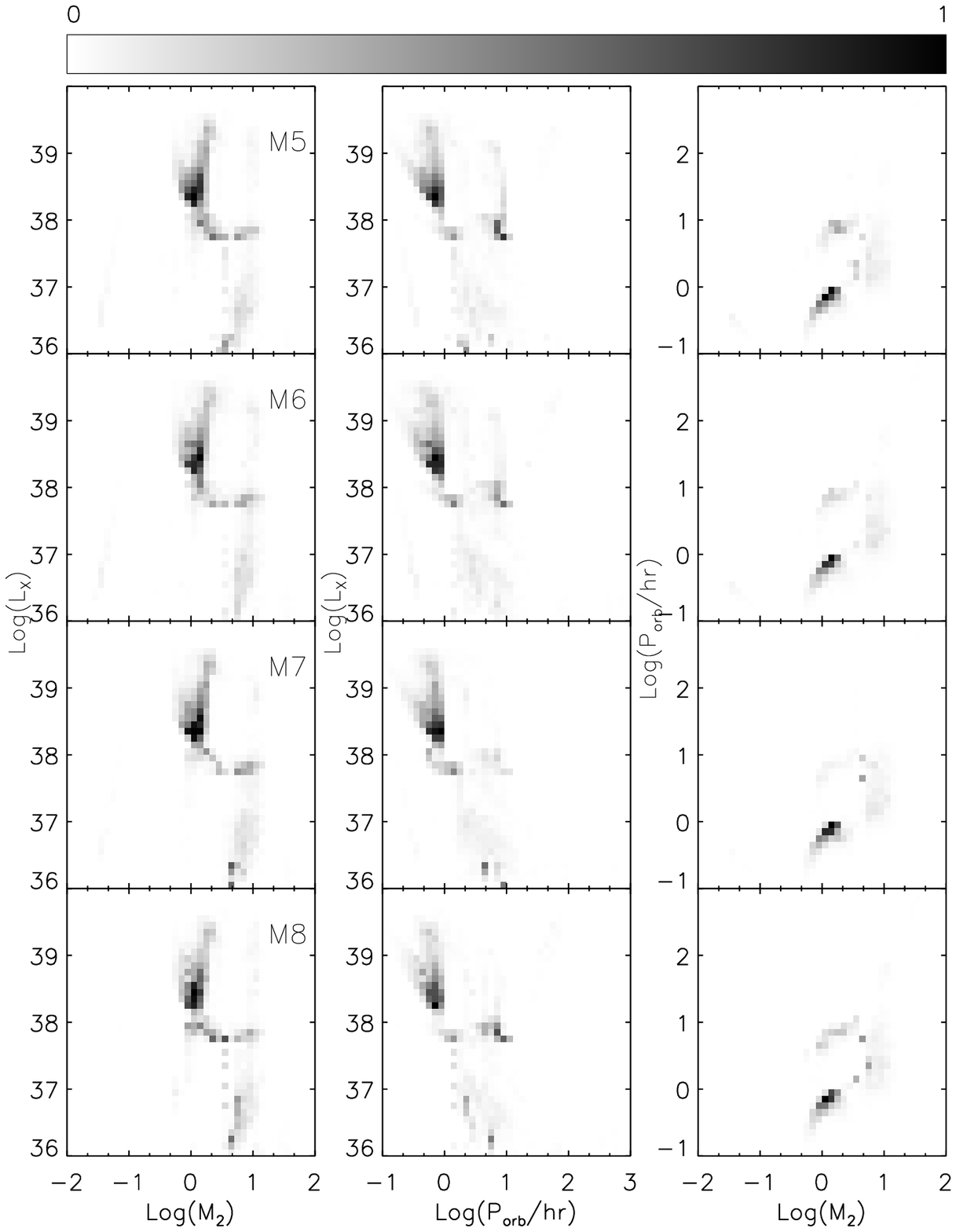}
\caption{Same as Fig.~5 but for sources in models M5-M8 from top to
bottom, respectively.}
  \label{Fig. 3b}
\end{figure}

\begin{figure}
  \centering
   \includegraphics[width=0.8\linewidth]{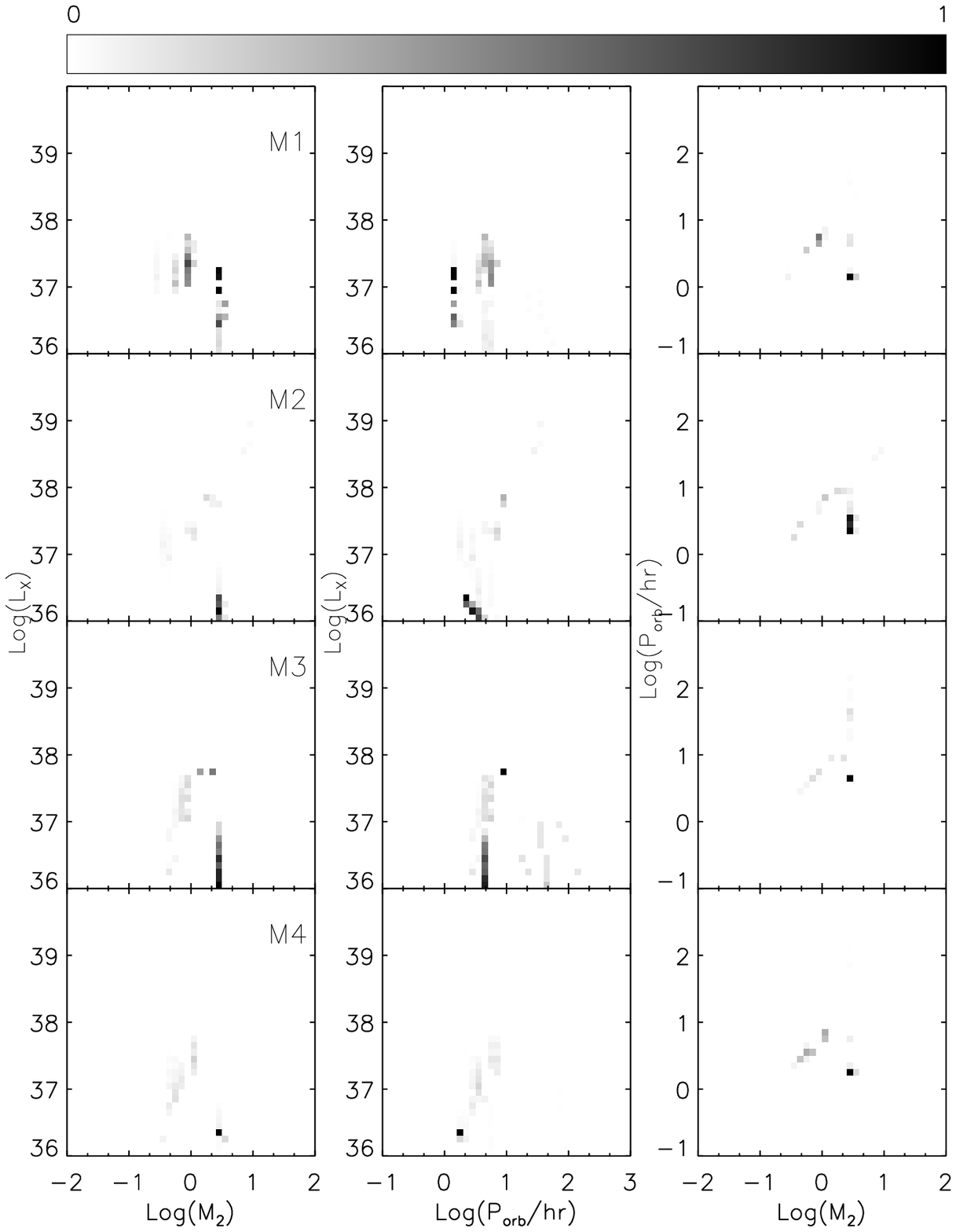}
\caption{The $L_{\rm X}-M_2$, $L_{\rm X}-P_{\rm orb}$, $P_{\rm
orb}-M_2$ distributions in the $300<R<1000$ pc region from left to
right, respectively. From top to bottom are models M5-M8,
respectively.}
  \label{Fig. 4a}
\end{figure}

\begin{figure}
  \centering
   \includegraphics[width=0.8\linewidth]{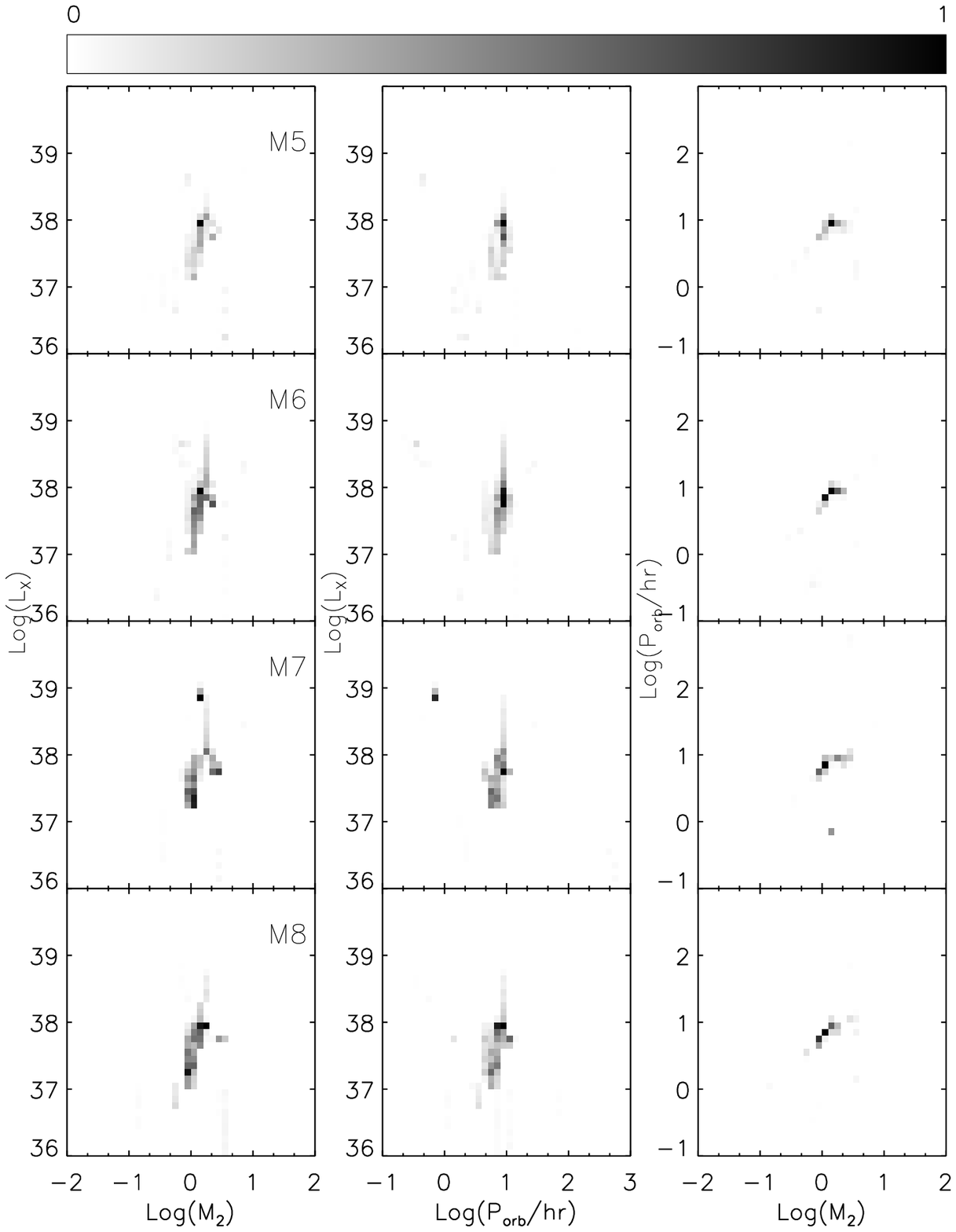}
\caption{Same as Fig.~7 but for sources in models M5-M8 from top to
bottom, respectively.}
  \label{Fig. 4b}
\end{figure}

\begin{figure}
  \centering
   \includegraphics[width=0.8\linewidth]{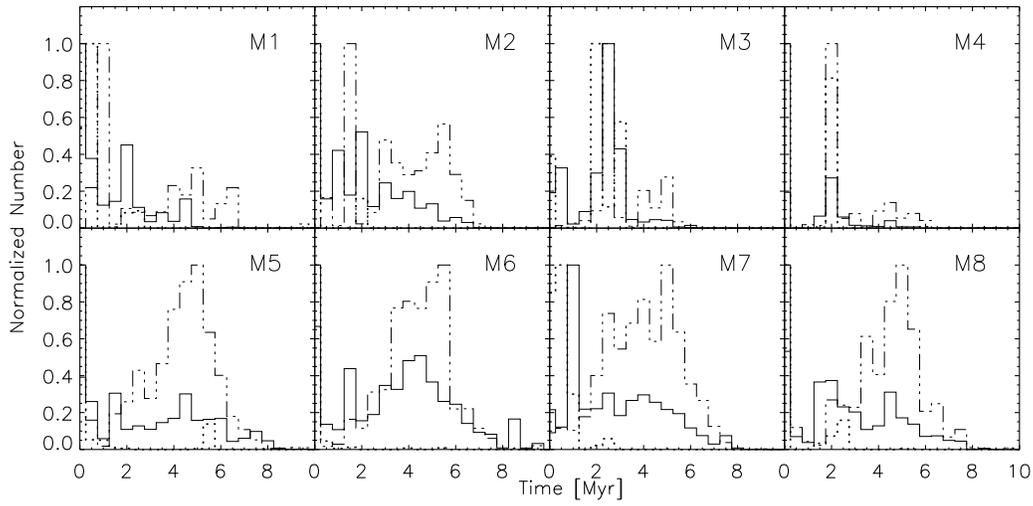}
\caption{The delay time distributions between SN and the beginning
of RLOF for sources in regions A (dash-dot-dotted line), B (solid
line) and C (dotted line), respectively. From left to right are
models M1-M4 (upper panel) and M5-M8 (bottom panel),
respectively.}
  \label{Fig. 5a}
\end{figure}

\begin{figure}
  \centering
   \includegraphics[width=0.8\linewidth]{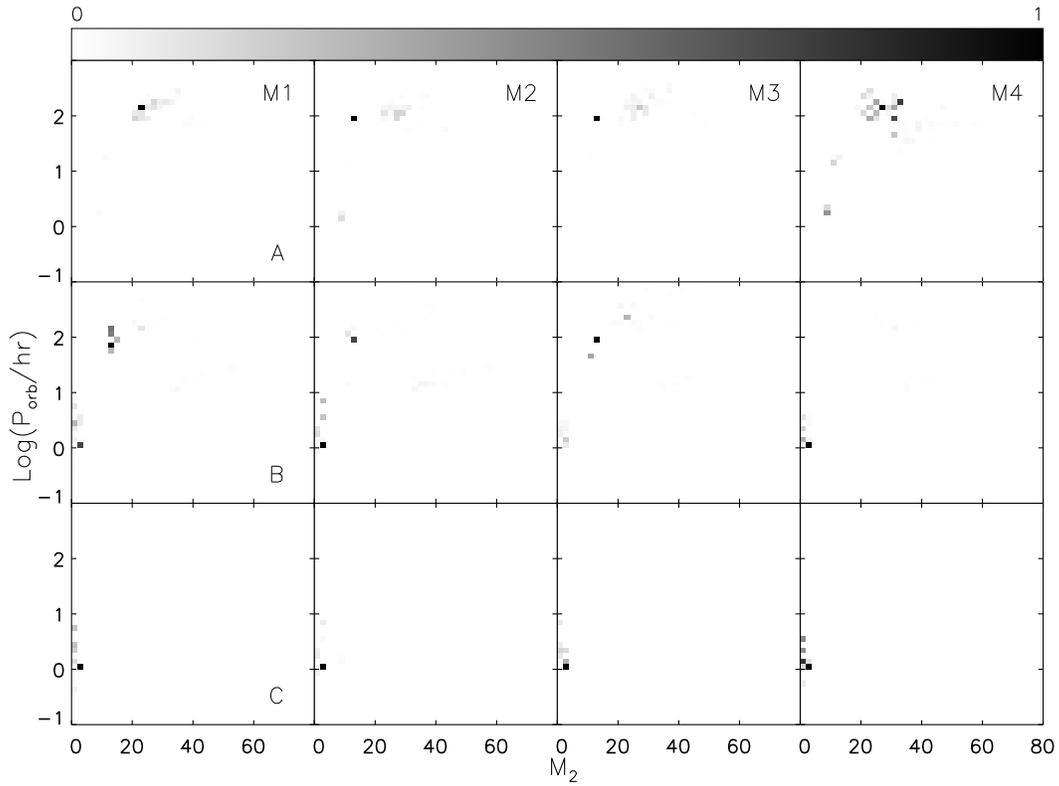}
\caption{The $P_{\rm orb, SNe}-M_{\rm 2, SNe}$ distributions in
regions A, B and C, respectively. From left to right are models
M1-M4, respectively.}
  \label{Fig. 5a}
\end{figure}

\begin{figure}
  \centering
   \includegraphics[width=0.8\linewidth]{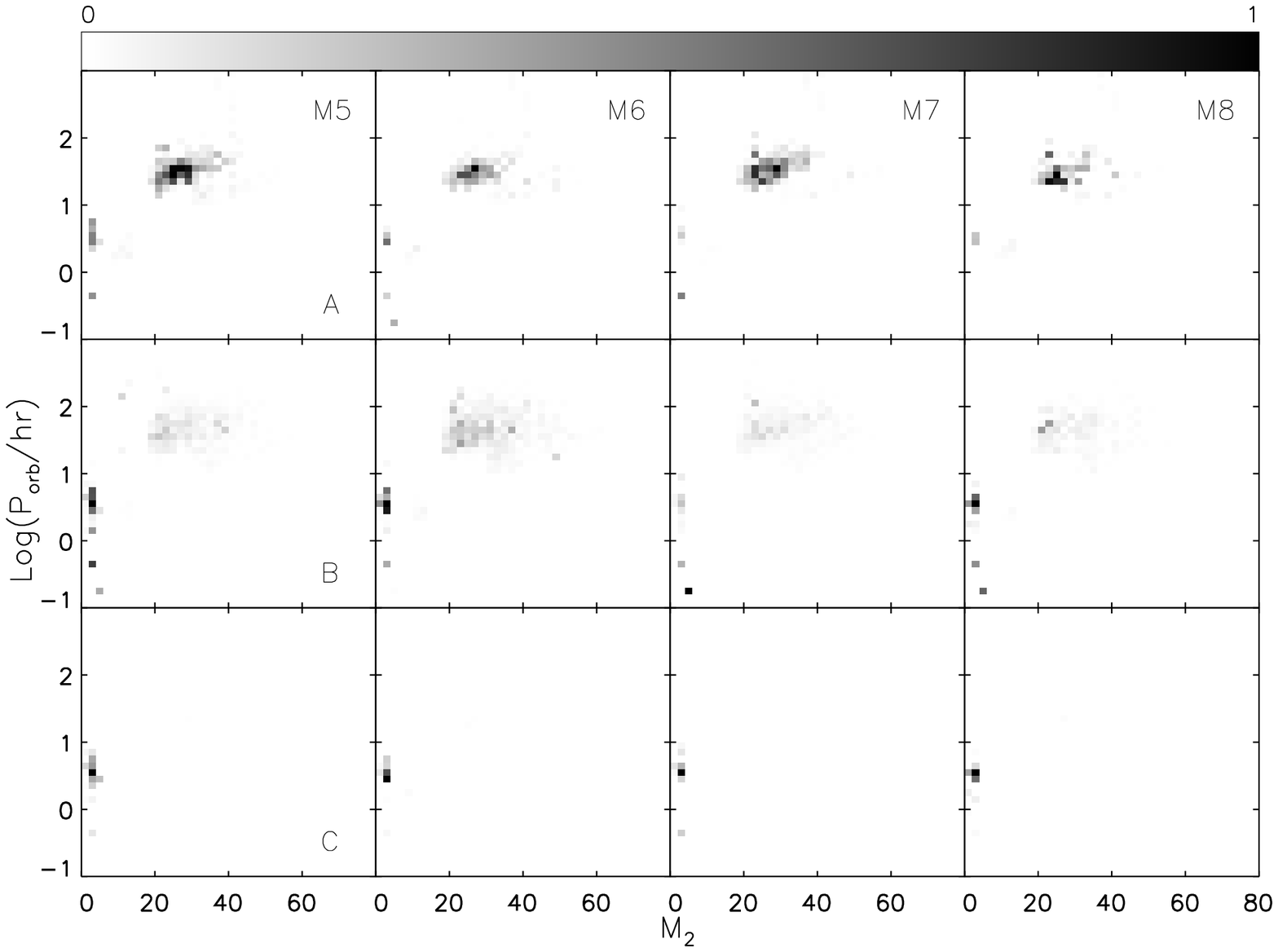}
\caption{Same as Fig.~10 but for sources in models M5-M8 from left
to right, respectively.}
  \label{Fig. 5b}
\end{figure}

\begin{figure}
  \centering
   \includegraphics[width=0.8\linewidth]{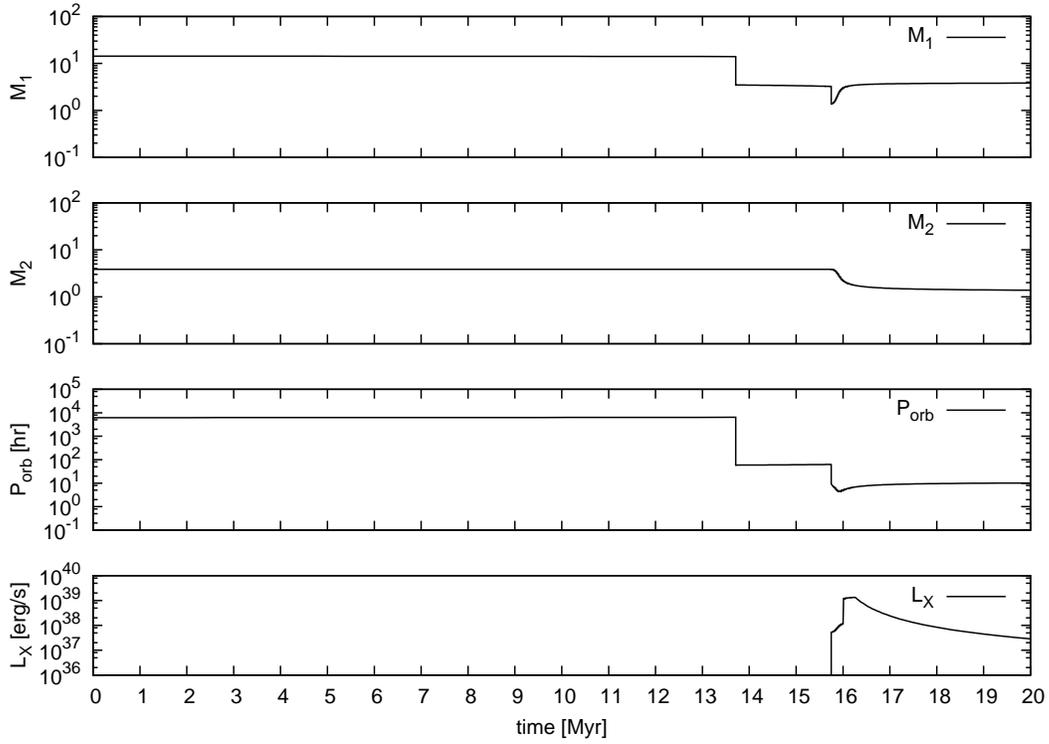}
\caption{The evolution of $M_1$, $M_2$, $P_{\rm orb}$, and $L_X$ for
an example XRB in region C.}
  \label{Fig. 6}
\end{figure}

\begin{figure}
  \centering
   \includegraphics[width=0.8\linewidth]{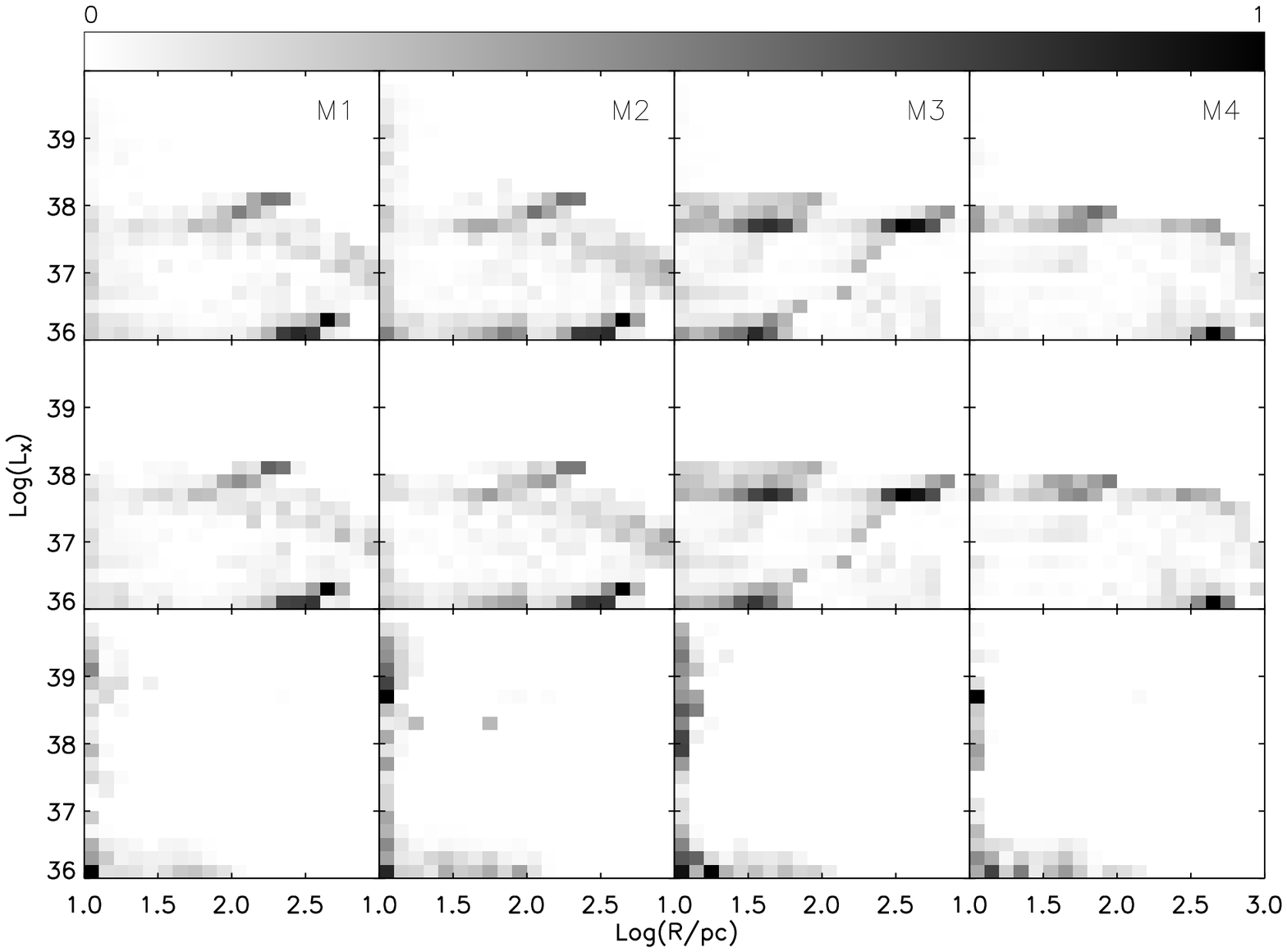}
\caption{Same as Fig.~3 but without AIC of accreting NSs.}
  \label{Fig. 7a}
\end{figure}

\begin{figure}
  \centering
   \includegraphics[width=0.8\linewidth]{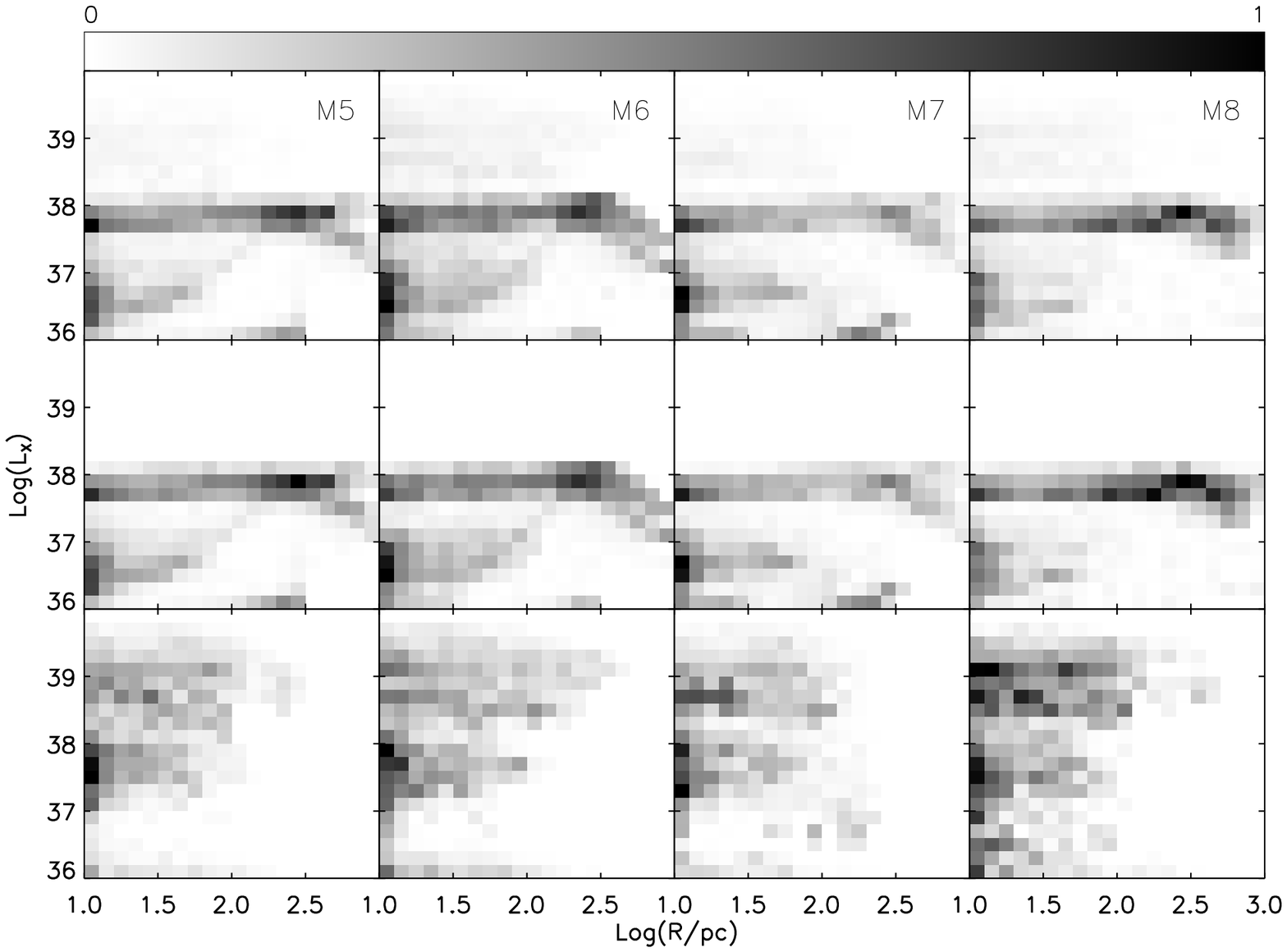}
\caption{Same as Fig.~4 but without AIC of accreting NSs..}
  \label{Fig. 7b}
\end{figure}

\begin{figure}
  \centering
   \includegraphics[width=0.8\linewidth]{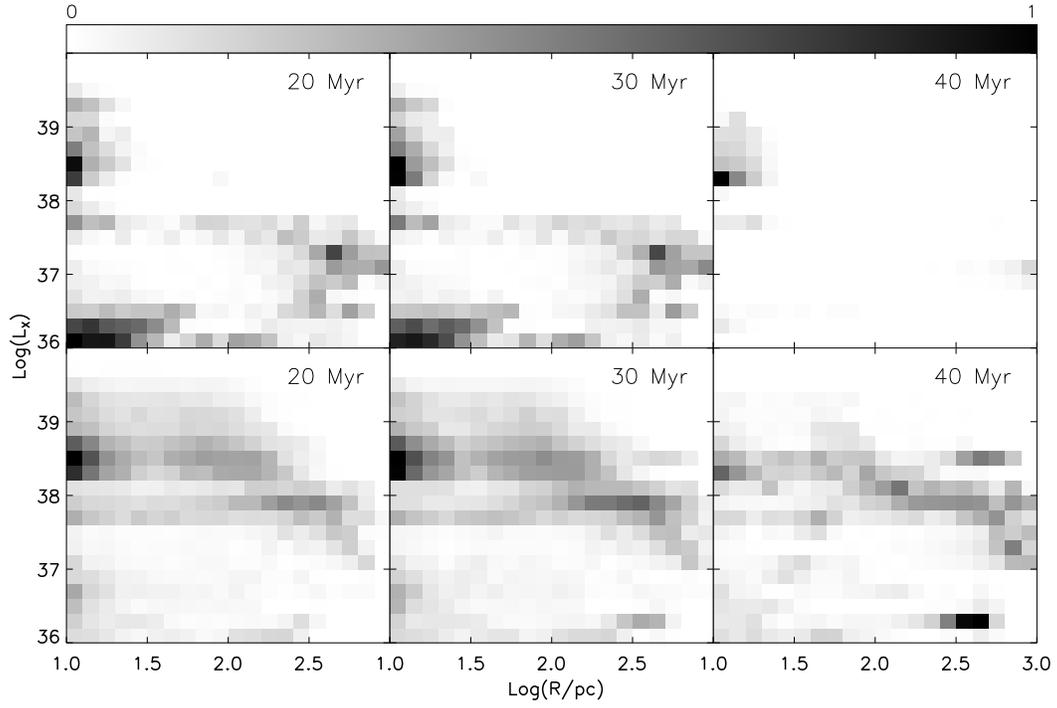}
\caption{The evolution of the $L_{\rm X}-R$ distribution for models
M1 (top panel) and M5 (bottom panel), respectively. From left to
right are for the ages of 20 Myr (left), 30 Myr (middle) and 40 Myr
(right), respectively.}
  \label{Fig. 8}
\end{figure}


\begin{thebibliography}{99}

\bibitem[\protect\citeauthoryear{Abt}{1983}]{abt83} Abt H. A., 1983, ARA\&A, 21, 343
\bibitem[\protect\citeauthoryear{Bailyn \& Grindlay}{1987}]{bailyn87} Bailyn C. D., Grindlay J. E., 1987, ApJ, 316, L25
\bibitem[\protect\citeauthoryear{Ballero et al.}{2007}]{ballero07} Ballero S., Matteucci F, Origlia L., Rich R., 2007, A\&A, 467, 123
%\bibitem[\protect\citeauthoryear{Barbosa \& Figer}{2004}]{barbosa04} Barbosa C., Figer D., 2004, Top 10 Problems on Massive Stars. arXiv:astro-ph/0408491
\bibitem[\protect\citeauthoryear{Bassa et al.}{2006}]{bassa06} Bassa C. G., van Kerkwijk M. H., Koester D., Verbunt F., 2006, A\&A, 456, 295
\bibitem[\protect\citeauthoryear{Belczynski}{2002}]{belczynski02} Belczynski K., Kalogera V., Bulik T., 2002, ApJ, 572, 407
\bibitem[\protect\citeauthoryear{Belczynski \& Taam}{2003}]{bel03} Belczynski K., Taam R. E., 2003, ApJ, 616, 1159
\bibitem[\protect\citeauthoryear{Chen et al.}{1997}]{chen97} Chen W., Shrader C. R., Livio M., 1997, ApJ, 491, 312
\bibitem[\protect\citeauthoryear{Dalton \& Sarazin}{1995}]{dalton95} Dalton W. W., Sarazin C. L., 1995, ApJ, 448, 369
\bibitem[\protect\citeauthoryear{Davies \& Hansen}{1998}]{davies98} Davies M. B., Hansen B. M. S., 1998, MNRAS, 301, 15
\bibitem[\protect\citeauthoryear{Dewi \& Tauris}{2000}]{dewi00} Dewi J.D.M., Tauris T.M., 2000, A\&A, 360, 1043
\bibitem[\protect\citeauthoryear{Fabbiano}{1989}]{fabb89} Fabbiano G., 1989, ARA\&A, 27, 87
\bibitem[\protect\citeauthoryear{Fryer et al.}{1999}]{fryer99} Fryer C. L., Woosley S. E., Hartmann D. H., 1999, ApJ, 526, 152
\bibitem[\protect\citeauthoryear{Fryer}{1999}]{fryer99a} Fryer C. L., 1999, ApJ, 522, 413
\bibitem[\protect\citeauthoryear{Fryer \& Kalogera}{2001}]{fryer01} Fryer C., Kalogera V., 2001, ApJ, 554, 548
\bibitem[\protect\citeauthoryear{Ho \& Filippenko}{1996a}]{ho96a} Ho L. C., Filippenko A. V., 1996a, ApJ, 466, L83
\bibitem[\protect\citeauthoryear{Ho \& Filippenko}{1996b}]{ho96b} Ho L. C., Filippenko A. V., 1996b, ApJ, 472, 600
\bibitem[\protect\citeauthoryear{Han et al.}{1995}]{han95} Han Z., Podsiadlowski P., Eggleton P. P., 1995, MNRAS, 270, 121
\bibitem[\protect\citeauthoryear{Hansen et al.}{1997}]{hansen97} Hansen B., Phinney E., 1997, MNRAS, 291, 569
\bibitem[\protect\citeauthoryear{Hobbs et al.}{2005}]{Hobbs} Hobbs G., Lorimer D. R., Lyne A. G., Kramer M., 2005, MNRAS, 360, 963
\bibitem[\protect\citeauthoryear{Hogeveen}{1990}]{hogeveen90} Hogeveen S. J., 1990, Ap\&SS, 173, 315
\bibitem[\protect\citeauthoryear{Hurley et al.}{2000}]{Hurley00} Hurley J. R., Pols O. R., Tout C. A., 2000, MNRAS, 315, 543
\bibitem[\protect\citeauthoryear{Hurley et al.}{2002}]{Hurley02} Hurley J. R., Tout C. A., Pols O. R., 2002, MNRAS, 329, 897
\bibitem[\protect\citeauthoryear{Hut \& Verbunt}{1983}]{hut83} Hut P., Verbunt F., 1983, Nature, 301, 587
\bibitem[\protect\citeauthoryear{Iben \& Tutukov}{1989}]{iben89} Iben I. Jr., Tutukov A. V., 1989, ApJ, 342, 430
\bibitem[\protect\citeauthoryear{Iben \& Livio}{1993}]{iben93} Iben I. Jr, Livio M., 1993, PASP, 105, 1373
%\bibitem[\protect\citeauthoryear{Iben et al.}{1995}]{iben95} Iben, I., Jr., Tutukov, A. V., Yungelon, L. R., 1995, ApJS, 100, 233
\bibitem[\protect\citeauthoryear{Garcia et al.}{2003}]{Garcia03} Garcia M. R., Miller J. M., McClintock J. E., King A. R., Orosz J.,
 2003, ApJ, 591, 388
% \bibitem[\protect\citeauthoryear{Grimm et al.}{2002}]{grimm02} Grimm H., Gilfanov M., Sunyaev R., 2002, A\&A, 391, 923
 \bibitem[\protect\citeauthoryear{Ivanova \& Kalogera}{2006}]{ivanova06} Ivanova N., Kalogera V., 2006, ApJ, 636, 985
% \bibitem[\protect\citeauthoryear{Jonker \& Nelemans}{2004}]{jonker04} Jonker P.G., Nelemans G., 2004, MNRAS, 354, 355
\bibitem[\protect\citeauthoryear{Kaaret et al.}{2003}]{kaaret03} Kaaret P., Corbel S., Prestwich A. H., Zezas A., 2003, Science, 299, 365
 \bibitem[\protect\citeauthoryear{Kaaret et al.}{2004}]{kaaret04} Kaaret P., Alonso-Herrero A., Gallagher J. S., Fabbiano G., Zezas A., Rieke M. J., 2004, MNRAS, 348, L28
\bibitem[\protect\citeauthoryear{Kalogera \& Webbink}{1998}]{kalogera98} Kalogera V.,  Webbink R. F., 1998, ApJ, 493, 351
\bibitem[\protect\citeauthoryear{Kalogera}{1999}]{kalogera99} Kalogera V., 1999, ApJ, 521, 723
\bibitem[\protect\citeauthoryear{Kaper \& van der Meer}{2007}]{Kaper07} Kaper L., van der Meer A., Massive Stars in Interactive Binaries, ASP Conference Series
367. Edited by Nicole St.-Louis and Anthony F.J. Moffat., San
Francisco: Astronomical Society of the Pacific, 2007., p.447
\bibitem[\protect\citeauthoryear{Kiel \& Hurley}{2006}]{kiel06} Kiel P. D., Hurley J. R., 2006, MNRAS, 369, 1152
\bibitem[\protect\citeauthoryear{Kiel \& Hurley}{2009}]{kiel09} Kiel P. D., Hurley J. R., 2009, MNRAS, 395, 2326
\bibitem[\protect\citeauthoryear{King et al.}{2001}]{king01} King A. R., Davies M. B., Ward M. J., Fabbiano G., Elvis M., 2001, ApJ, 552, L109
\bibitem[\protect\citeauthoryear{Kobulnicky \& Fryer}{2007}]{kobulnicky07} Kobulnicky H. A., Fryer C. L., 2007, ApJ, 670, 747
\bibitem[\protect\citeauthoryear{K\"{o}rding et al.}{2002}]{kording02} K\"{o}rding E., Falcke H., Markoff S., 2002, A\&A, 382, L13
\bibitem[\protect\citeauthoryear{Kroupa et al.}{1993}]{Kroupa} Kroupa P., Tout C. A., Gilmore G., 1993, MNRAS, 262, 545
\bibitem[\protect\citeauthoryear{Kulkarni et al.}{1993}]{kulkarni93} Kulkarni S. R., Hut P., McMillian S., 1993, Nature, 364, 421
\bibitem[\protect\citeauthoryear{Larson}{2001}]{larson01} Larson R. B., 2001, in IAU Symp. 200, The Formation of Binary
Stars, ed. H. Zinnecker \& R. D. Mathieu (San Francisco: ASP), 93
%\bibitem[\protect\citeauthoryear{Lee et al.}{2002}]{lee02} Lee, C. H., Brown, G. E., Wijers, R.A.M.J., 2002, ApJ, 575, L996
\bibitem[\protect\citeauthoryear{Liu \& Li}{2007}]{liu07} Liu X. W., Li X. D., 2007, ChJA\&A, 7, 389
%\bibitem[\protect\citeauthoryear{Liu et al.}{2006}]{liujf06} Liu J. F., Bregman J. N., Irwin J., 2006, ApJ, 642, 171
\bibitem[\protect\citeauthoryear{Livio \& Soker}{1988}]{livio88} Livio M., Soker N., 1988, ApJ, 329, 764
\bibitem[\protect\citeauthoryear{Lucy}{2006}]{lucy06} Lucy L. B., 2006, A\&A, 457, 629
\bibitem[\protect\citeauthoryear{Lyne \& Lorimer}{1994}]{lyne94} Lyne A. G., Lorimer D. R., 1994, Nature, 369, 124
\bibitem[\protect\citeauthoryear{Massey \& Hunter}{1998}]{massey98} Massey P. M., Hunter D. A., 1998, ApJ, 493, 180
\bibitem[\protect\citeauthoryear{McCrady et al.}{2003}]{mcCrady03} McCrady N., Gilbert A. M., Graham J. R., 2003, ApJ, 596, 240
\bibitem[\protect\citeauthoryear{Nelemans et al.}{1999}]{nelemans99} Nelemans G., Tauris T. M., van den Heuvel E. P. J., 1999, A\&A, 352, L87
%\bibitem[\protect\citeauthoryear{Mengel et al.}{2002}]{mengel02} Mengel S., Lehnert M. D., Thatte N., Genzel R., 2002, A\&A, 383, 137
\bibitem[\protect\citeauthoryear{Ostriker}{1975}]{ostriker75} Ostriker J. P., 1975, paper presented at IAU Symp., 73, The
Structure and Evolution of Close Binary Systems
\bibitem[\protect\citeauthoryear{Paczy\'{n}ski}{1976}]{paczynski76} Paczy\'{n}ski B., 1976, In: Eggleton P., Mitton S., Whelan J. (eds.) Structure
and Evolution in Close Binary Systems. Proc. IAU Symp. 73, Reidel,
Dordrecht, p. 75
\bibitem[\protect\citeauthoryear{Paczy\'{n}ski}{1990}]{paczynski90} Paczy\'{n}ski B., 1990, ApJ, 348, 485
\bibitem[\protect\citeauthoryear{Phinney \& Sigurdsson}{1991}]{phinney91} Phinney E. S., Sigurdsson S., 1991, Nature, 349, 220
\bibitem[\protect\citeauthoryear{Podsiadlowski et al.}{2003}]{podsiadlowski03} Podsiadlowski P., Rappaport S. A., Han Z., 2003, MNRAS, 341, 385
\bibitem[\protect\citeauthoryear{Portegies Zwart et al.}{1997}]{portegies97} Portegies Zwart, S. F., Verbunt, F., Ergma, E., 1997, A\&A, 321, 207
\bibitem[\protect\citeauthoryear{Portegies Zwart et al.}{1999}]{portegies99} Portegies Zwart S. F., Makino J., McMillian S. L. W., Hut P., 1999, A\&A, 348, 117
\bibitem[\protect\citeauthoryear{Pourbaix et al.}{2004}]{pourbaix04} Pourbaix D. et al., 2004, A\&A, 424, 727
\bibitem[\protect\citeauthoryear{Rappaport, Podsiadlowski \& Pfahl}{2004}]{rappaport04} Rappaport S. A., Podsiadlowski P., Pfahl E., 2004, MNRAS, 361, 971
%\bibitem[\protect\citeauthoryear{Reg\H{o}s \& Tout}{1995}]{roges95} Reg\H{o}s E., Tout C. A., 1995, MNRAS, 273, 146
\bibitem[\protect\citeauthoryear{Ricker \& Taam}{2008}]{ricker08} Ricker P. M., Taam R. E., 2008, ApJ, 672, L41
%\bibitem[\protect\citeauthoryear{Romani}{1992}]{romani92} Romani, R. W., 1992, ApJ, 399, 621
\bibitem[\protect\citeauthoryear{Sandquist et al.}{2000}]{sandquist00} Sandquist E. L., Taam R. E., Burkert A., 2000, ApJ, 533, 984
\bibitem[\protect\citeauthoryear{Sigurdsson \& Hernquist}{1993}]{sigurdsson93} Sigurdsson S., Hernquist L., 1993, Nature, 364, 423
\bibitem[\protect\citeauthoryear{Smith \& Gallagher}{2001}]{smith01} Smith L. J., Gallagher J. S., 2001, MNRAS, 326, 1027
\bibitem[\protect\citeauthoryear{Sternberg}{1998}]{sternberg98} Sternberg A., 1998, ApJ, 506, 721
%\bibitem[\protect\citeauthoryear{Swartz et al.}{2004}]{swartz04} Swartz D. A., Ghosh K. K., Tennant A. F., Wu K., 2004, ApJS, 154, 519
\bibitem[\protect\citeauthoryear{Taam \& Bodenheimer}{1989}]{Taam89} Taam R., Bodenheimer P., 1989, ApJ, 337, 849
\bibitem[\protect\citeauthoryear{Taam}{1994}]{taam94} Taam R. E., 1994, in ASP Conf. Ser., 56, Interacting Binary Stars, ed. A. W.
Shafter (San Francisco: ASP), 208
\bibitem[\protect\citeauthoryear{Taam \& Sandquist}{2000}]{taam00} Taam R., Sandquist E., 2000, ARA\&A, 38,113
\bibitem[\protect\citeauthoryear{Tauris \& van den Heuvel}{2006}]{tauris06} Tauris T.M., van den Heuvel E. P. J., 2006, in Compact Stellar X-ray Sources,
ed. W. H. G. Lewin \& M. van der Klis, Cambridge Univ. Press, (Cambridge), 623
%\bibitem[\protect\citeauthoryear{Temple et al.}{2005}]{temple05} Temple R., Raychaudhury S., Stevens I., 2005, MNRAS, 362, 581
%\bibitem[\protect\citeauthoryear{Terman et al.}{1998}]{terman98} Terman, J., Taam, R., Savage, C., 1998, MNRAS, 293, 113
\bibitem[\protect\citeauthoryear{Tutukov \& Yungelon}{1993}]{tutukov93} Tutukov A. V., Yungelon L. R., 1993, MNRAS, 260, 675
\bibitem[\protect\citeauthoryear{van den Heuvel}{1975}]{heuvel75} van den Heuvel E.P.J., 1975, ApJ, 198, L109
%\bibitem[\protect\citeauthoryear{van den Heuvel}{1994}]{heuvel94} van den Heuvel, E.P.J., 1994, in Interacting Binaries, Saas-Fee course 22, (Springer, Heidelberg)
%p. 263
%\bibitem[\protect\citeauthoryear{van den Heuvel}{1994}]{heuvel94aa} van den Heuvel E.P.J., 1994, A\&A 291, L39
%\bibitem[\protect\citeauthoryear{van den Heuvel}{1994}]{heuvel94} van den Heuvel, E.P.J., 1994, A\&A, 291, L39
\bibitem[\protect\citeauthoryear{van den Heuvel et al.}{2000}]{heuvel00} van den Heuvel E. P. J., Portegies Zwart S. F., Bhattacharya D.,
Kaper L., 2000, A\&A, 364, 563
\bibitem[\protect\citeauthoryear{Van Paradijs \& White}{1995}]{paradijs95}Van Paradijs J., White N., 1995, ApJ, 447, L33
\bibitem[\protect\citeauthoryear{Van Paradijs}{1996}]{paradijs96} Van Paradijs J., 1996, ApJ, 464, L139
\bibitem[\protect\citeauthoryear{Verbunt \& Hut}{1987}]{verbunt87} Verbunt F., Hut P., 1987, IAUS, 125, 187
%\bibitem[\protect\citeauthoryear{Verbunt \& Hut}{1987}]{verbunt87} Verbunt, F., Hut, P., 1987, in The Origin and Evolution of Neutron Stars,
%ed. D. J. Helfand \& J.-H. Huang (Dordrecht: Reidel), 187
\bibitem[\protect\citeauthoryear{Verbunt \& van den Heuvel}{1994}]{verbunt94} Verbunt F., van den Heuvel E., 1994, X-ray Binaries, Cambridge University Press, p. 457
\bibitem[\protect\citeauthoryear{Webbink et al.}{1983}]{webbink83} Webbink R. F., Rappaport S., Savonije G. J., 1983, ApJ, 270, 678
\bibitem[\protect\citeauthoryear{Webbink}{1984}]{webbink84} Webbink R.F., 1984, ApJ, 277, 355
\bibitem[\protect\citeauthoryear{Webbink}{1992}]{webbink92} Webbink R. F., 1992, in X-Ray Binaries and Recycled Pulsars, van den Heuvel E. P. J.,
Rappaport S. A., (NATO ASI Ser. C, 377; Dordrecht: Kluwer), 269
\bibitem[\protect\citeauthoryear{White \& Van Paradijs}{1996}]{white96} White N.E., Van Paradijs J., 1996, ApJ, 473, L25
\bibitem[\protect\citeauthoryear{Xu \& Li}{2010}]{xu10} Xu X.J., Li X. D., 2010, in preparation
\bibitem[\protect\citeauthoryear{Zuo et al.}{2008}]{zuo08} Zuo Z. Y., Li X. D., Liu X. W., 2008, MNRAS, 387, 121
\end{thebibliography}
\end{document}